\documentclass[11pt]{article}
\usepackage{amssymb,amsmath,amsfonts}
\usepackage{graphicx}
\usepackage{graphics}
\usepackage{eepic,epsfig}

\begin{document}

\title{Fermionic current from topology and boundaries with\\
applications to higher-dimensional models and nanophysics}
\author{ S. Bellucci$^{1}$\thanks{%
E-mail: bellucci@lnf.infn.it } and A. A. Saharian$^{2}$\thanks{%
E-mail: saharian@ysu.am } \vspace{0.3cm}\\
\textit{$^1$ INFN, Laboratori Nazionali di Frascati,}\\
\textit{Via Enrico Fermi 40, 00044 Frascati, Italy} \vspace{0.3cm}\\
\textit{$^2$ Department of Physics, Yerevan State University,}\\
\textit{1 Alex Manoogian Street, 0025 Yerevan, Armenia }}
\maketitle

\begin{abstract}
We investigate combined effects of topology and boundaries on the
vacuum expectation value (VEV) of the fermionic current in the
space with an arbitrary number of toroidally compactified
dimensions. As a geometry of boundaries we consider two parallel
plates on which the fermion field obeys bag boundary conditions.
Along the compact dimensions, periodicity conditions are imposed
with arbitrary phases. In addition, the presence of a constant
gauge field is assumed. The nontrivial topology gives rise to an
Aharonov-Bohm effect for the fermionic current induced by the
gauge field. It is shown that the VEV of the charge density
vanishes and the current density has nonzero expectation values
for the components along compact dimensions only. The latter are
periodic odd functions of the magnetic flux with the period equal
to the flux quantum. In the region between the plates, the VEV of
the fermionic current is decomposed into pure topological, single
plate and interference parts. For a massless field the single
plate part vanishes and the interference part is distributed
uniformly. The corresponding results are generalized for
conformally-flat spacetimes. Applications of the general formulas
to finite-length carbon nanotubes are given within the framework
of the Dirac model for quasiparticles in graphene. In the absence
of the magnetic flux, two sublattices of the honeycomb graphene
lattice yield opposite contributions and the fermionic current
vanishes. A magnetic flux through the cross section of the
nanotube breaks the symmetry allowing the current to flow along
the compact dimension.
\end{abstract}

\bigskip


\bigskip

\section{Introduction}

\label{sec:Introd}

In quantum field theory, the boundary conditions imposed on the field
operator give rise to a number of interesting physical effects. In the first
class of models, these conditions are due to presence of boundaries having
different physical nature, like macroscopic bodies in QED, extended
topological defects, horizons and so on. In this type of problems the field
operator obeys the boundary condition on some space-like surfaces. A
well-known quantum effect induced by this kind of boundary conditions is the
Casimir effect (for reviews see Ref. \cite{Most97}). The Casimir effect is
among the most striking macroscopic manifestations of nontrivial properties
of the quantum vacuum. The boundary conditions imposed on the field operator
alter the zero-point modes of a quantized field and shift the vacuum
expectation values (VEVs) of quantities, such as the energy density and
stresses. This leads to the appearance of forces acting on constraining
boundaries.

In the second class of models, the boundary conditions arise due
to the nontrivial topology of the space. The latter induces
periodicity conditions imposed on the field operator along compact
dimensions. This type of models appear in many high-energy
theories of fundamental physics, including supergravity and
superstring theories. Models of a compact universe with nontrivial
topology may also play an important role by providing proper
initial conditions for inflation in the early stages of the
Universe expansion \cite{Lind04n}. An interesting application of
the field theoretical models with compact dimensions appeared in
nanophysics recently. For long-wavelengths, the dynamics of
quasiparticles in a graphene sheet is well described in terms of
the Dirac-like theory in 2-dimensional space with
the Fermi velocity playing the role of the speed of light (see Ref.~\cite{Cast09}%
). In the geometry of a single-walled carbon nanotube, which is
generated by rolling up a graphene sheet to form a cylinder, the
background space for the corresponding Dirac model has topology
$R^{1}\times S^{1}$. For toroidal carbon nanotubes, an additional
compactification along the tube axis leads to the topology
$(S^{1})^{2}$.

In models with nontrivial topology, the periodicity conditions
along compact dimensions give rise to Casimir-type contributions
in the VEVs of various physical observables. In Kaluza-Klein
models, the topological Casimir effect has been used for the
stabilization of moduli fields and as a source for dynamical
compactification of extra dimensions. As it has been discussed in
Refs.~\cite{Eliz01}, the Casimir energy related to the compact
subspace of extra dimensions can serve as a model for dark energy
driving the accelerated expansion of the universe at the present
epoch. Note that, the recent measurements of the Casimir forces
between macroscopic bodies provide a sensitive test to constrain
the parameters of various types of long-range interactions
predicted by unification theories \cite{Most87}. The influence of
extra compactified dimensions on the Casimir effect in the
geometry of two parallel plates has been recently discussed for
scalar \cite{Chen06}, electromagnetic \cite{Popp04}, and Dirac
fermion \cite{Bell09b,Eliz11} fields. The combined quantum vacuum
effects from boundaries (branes) and from the non-trivial topology
of spatial dimensions in braneworld models on anti-de Sitter bulk
are considered in Ref. \cite{Flac03} (for the Casimir energy and
stresses in braneworlds see Ref.~\cite{Gold00}). In these models
the Casimir forces provide a natural mechanism for stabilizing the
radion field, as required for a complete solution of the hierarchy
problem between the gravitational and electroweak mass scales.

In the papers cited above, the vacuum energy and the forces acting
on the constraining boundaries were considered. For charged fields
an additional important characteristic of the vacuum state,
bilinear in the field, is the VEV of the current density. In
Ref.~\cite{Bell10}, the VEV of the current density is evaluated
for a massive fermionic field in spaces with toroidally
compactified dimensions. It has been shown that the nontrivial
topology of the background spacetime leads to the Aharonov-Bohm
effect for the fermionic current induced by the gauge field.
Continuing in this line of investigation, in the present paper we
consider the combined effects of boundaries and compact spatial
dimensions on the VEV of the fermionic current. Although the
corresponding operator is local, due to the global nature of the
vacuum, this quantity carries an important information about the
topology of the background space. The current acts as the source
in the Maxwell equations and therefore plays an important role in
modelling a self-consistent dynamics involving the electromagnetic
field. As the boundary geometry here we will consider two parallel
plates on which the fermionic field obeys the MIT bag boundary
condition. In addition, we assume the presence of a constant gauge
field (the fermionic current in a conical space with a circular
boundary has been recently discussed in Ref.~\cite{Beze10}).

The outline of the paper is as follows. In the next section, we specify the
mode functions and the eigenvalues of the momentum for the Dirac equation in
the region between two plates with bag boundary conditions on them. The VEV
of the fermionic current in this region is investigated in Sect.~\ref{sec:FC}%
. This VEV is decomposed into a pure topological, single plate and
interference parts. We also give a generalization for
conformally-flat
background spacetimes with toroidally compact dimensions. In Sect.~\ref%
{sec:Nano} we apply the general formulas for the investigation of the
fermionic current in finite-length metallic and semiconductor carbon
nanotubes, within the framework of the Dirac model for quasiparticles in
graphene. The main results of the paper are summarized in Sect.~\ref%
{sec:Conc}. In Appendix we derive an alternative expression for
the VEV of the fermionic current in the geometry of a single
plate.

\section{Mode functions}

\label{sec:Modes}

We consider a Dirac fermion field $\psi (x)$ in background of a $(D+1)$%
-dimensional flat spacetime, in the presence of a gauge field
$A_{\mu }$. The dynamics of the field is described by the Dirac
equation
\begin{equation}
(i\gamma ^{\mu }D_{\mu }-m)\psi (x)=0\ ,  \label{Direq}
\end{equation}%
with $D_{\mu }=\partial _{\mu }+ieA_{\mu }$. The dimensions for Dirac
matrices $\gamma ^{\mu }=(\gamma ^{0},\boldsymbol{\gamma })$ are $%
N_{D}\times N_{D}$, where $N_{D}=2^{[(D+1)/2]}$ and the square
brackets denote the integer part of $(D+1)/2$. In the discussion
below we use the Dirac representation:
\begin{equation}
\gamma ^{0}=\left(
\begin{array}{cc}
1 & 0 \\
0 & -1%
\end{array}%
\right) ,\;\boldsymbol{\gamma }=\left(
\begin{array}{cc}
0 & \boldsymbol{\sigma } \\
-\boldsymbol{\sigma }^{+} & 0%
\end{array}%
\right) ,  \label{DiracMat}
\end{equation}%
with $\boldsymbol{\sigma }=(\sigma _{1},\ldots ,\sigma _{D})$. For the
matrices $\sigma _{\mu }$ in Eq. (\ref{DiracMat}) we have the
anticommutation relations $\sigma _{\mu }\sigma _{\nu }^{+}+\sigma _{\nu
}\sigma _{\mu }^{+}=2\delta _{\mu \nu }$. In a two-dimensional space the
irreducible representation corresponds to $N_{D}=2$ and the corresponding
Dirac matrices can be taken as $\gamma ^{\mu }=(\sigma _{\text{\textrm{P}}%
3},i\sigma _{\text{\textrm{P}}1},i\sigma _{\text{\textrm{P}}2})$, where $%
\sigma _{\text{\textrm{P}}\mu }$ are the Pauli matrices.

We assume that $q=D-p-1$ spatial dimensions with the Cartesian coordinates $%
\mathbf{z}_{q}=(z_{p+2},\ldots ,z_{D})$ are toroidally compactified. The
length of the $l$-th compact dimension will be denoted by $L_{l}$, so $%
0\leqslant z_{l}\leqslant L_{l}$ for $l=p+2,\ldots ,D$. The
remaining coordinates $\mathbf{z}_{p+1}=(z_{1},\ldots
,z_{p+1}\equiv z)$ are not compactified with $-\infty
<z_{l}<\infty $, $l=1,\ldots ,p+1$. Hence, we consider the
background space with topology $R^{p+1}\times (S^{1})^{q}$. Along
the compact dimensions we impose on the field operator
quasiperiodic boundary conditions:%
\begin{equation}
\psi (t,\mathbf{z}_{p+1},\mathbf{z}_{q}+L_{l}\mathbf{e}_{l})=e^{2\pi i\alpha
_{l}}\psi (t,\mathbf{z}_{p+1},\mathbf{z}_{q}),  \label{BC}
\end{equation}%
with constant phases $|\alpha _{l}|\leqslant 1/2$ and with $\mathbf{e}_{l}$\
being the unit vector along the direction of the coordinate $z_{l}$, $%
l=p+2,\ldots ,D$. For untwisted and twisted fermionic fields one has special
values $\alpha _{l}=0$ and $\alpha _{l}=1/2$, respectively.

Our main interest in this paper is the VEV of the fermionic current $j^{\mu
}=\bar{\psi}\gamma ^{\mu }\psi $, where $\bar{\psi}=\psi ^{\dagger }\gamma
^{0}$ is the Dirac adjoint and, as usual, the dagger denotes Hermitian
conjugation. We assume the presence of two parallel plates placed at $z=0$
and $z=a$, on which the field operator obeys MIT bag boundary conditions
\begin{equation}
\left( 1+i\gamma ^{\mu }n_{\mu }\right) \psi (x)=0\ ,\quad z=0,a,
\label{BagCond}
\end{equation}%
where $n_{\mu }$ is the outward oriented normal to the boundary. From these
conditions it follows that on the boundaries $n_{\mu }j^{\mu }=0$, i.e. the
component of the current along the normal to the boundaries vanishes on the
plates. In what follows we will consider the region between the plates, $%
0<z<a$, with $n_{\mu }=-\delta _{\mu }^{p+1}$ and $n_{\mu }=\delta _{\mu
}^{p+1}$ for the plates at $z=0$ and $z=a$, respectively. The expressions
for the VEVs in the regions $z<0$ and $z>a$ are obtained by the limiting
transition.

The VEV of the fermionic current is expressed in terms of the two-point
function $S_{rs}^{(1)}(x,x^{\prime })=\langle 0|[\psi _{r}(x),\bar{\psi}%
_{s}(x^{\prime })]|0\rangle $, where $r$, $s$ are spinor indices and $%
|0\rangle $ stands for the vacuum state. The expression for the VEV reads:%
\begin{equation}
\langle j^{\mu }(x)\rangle \equiv \langle 0|j^{\mu }(x)|0\rangle =-\frac{1}{2%
}\mathrm{Tr}(\gamma ^{\mu }S^{(1)}(x,x)).  \label{VEVj}
\end{equation}%
Let $\{\psi _{\beta }^{(+)}(x),\psi _{\beta }^{(-)}(x)\}$ be a complete set
of positive- and negative-energy solutions to the Dirac equation with a set
of quantum numbers $\beta $ specifying the modes (see below). Expanding the
fermionic field operator in terms of these solutions with the coefficients
being the annihilation and creation operators, the VEV of the current is
presented in the form of the mode sum
\begin{equation}
\langle j^{\mu }\rangle =\frac{1}{2}\sum_{\beta }[\bar{\psi}_{\beta
}^{(-)}(x)\gamma ^{\mu }\psi _{\beta }^{(-)}(x)-\bar{\psi}_{\beta
}^{(+)}(x)\gamma ^{\mu }\psi _{\beta }^{(+)}(x)].  \label{VEVj1}
\end{equation}%
This expression is divergent, hence some regularization and
subsequent renormalization procedure is necessary. The important
point here is that, owing to the flatness of the background
spacetime, the structure of divergences is the same as for the
topologically trivial Minkowski spacetime. As a result, the
renormalization is reduced to the subtraction from the VEV of the
corresponding Minkowskian quantity. In what follows, in order to
make the expression on the right-hand side of Eq. (\ref{VEVj1})
finite, we will assume the presence of some cutoff function,
without writing it explicitly. The special form of the latter will
not be important in the remainder of the discussion.

For the evaluation of the VEV by Eq. (\ref{VEVj1}) we need to have the mode
functions $\psi _{\beta }^{(\pm )}(x)$ for the equation (\ref{Direq})
obeying the periodicity conditions (\ref{BC}) and the boundary conditions (%
\ref{BagCond}). For the gauge field we assume that $A_{\mu }=\mathrm{const}$%
. Although the corresponding field strength vanishes, the
nontrivial topology of the space leads to the Aharonov-Bohm-like
effect for the fermionic current. In the problem under
consideration, the mode sum (\ref{VEVj1}) involves the integration
over the momentum along uncompactified dimensions and the
components of the constant gauge field along these dimensions are
simply removed by shifting the integration variable. Consequently,
the VEV does not depend on these components. For that reason we
will assume a nonzero vector
potential along compact dimensions only: $A_{\mu }=(0,-\mathbf{A})$ with $%
\mathbf{A}=(0,\mathbf{A}_{q})$ and $\mathbf{A}_{q}=(A_{p+2},\ldots ,A_{D})$.

With this choice, the mode functions are obtained from those discussed in
Ref.~\cite{Eliz11} by a simple generalization:%
\begin{eqnarray}
\psi _{\beta }^{(+)} &=&A_{\beta }^{(+)}e^{-i\omega ^{(+)}t}\left(
\begin{array}{c}
\varphi ^{(+)} \\
\frac{-i}{\omega ^{(+)}+m}\boldsymbol{\sigma }^{+}\cdot (\boldsymbol{\nabla }%
-ie\mathbf{A})\varphi ^{(+)}%
\end{array}%
\right) ,  \notag \\
\psi _{\beta }^{(-)} &=&A_{\beta }^{(-)}e^{i\omega ^{(-)}t}\left(
\begin{array}{c}
\frac{i}{\omega ^{(-)}+m}\boldsymbol{\sigma }\cdot (\boldsymbol{\nabla }-ie%
\mathbf{A})\varphi ^{(-)} \\
\varphi ^{(-)}%
\end{array}%
\right) ,  \label{modes}
\end{eqnarray}%
For the upper and lower components of the positive- and negative-energy
functions one has the expressions%
\begin{equation}
\varphi ^{(\pm )}=e^{\pm i\mathbf{k}_{\parallel }^{(\pm )}\cdot \mathbf{z}%
_{\parallel }}(\varphi _{+}^{(\pm )}e^{ik_{p+1}z}+\varphi _{-}^{(\pm
)}e^{-ik_{p+1}z}),  \label{phixi1}
\end{equation}%
with $\mathbf{z}_{\parallel }=$ $(z_{1},\ldots ,z_{p},z_{p+2},\ldots ,z_{D})$
being the coordinates parallel to the plates. For the corresponding momentum
we have $\mathbf{k}_{\parallel }^{(\pm )}=(\mathbf{k}_{p},\mathbf{k}%
_{q}^{(\pm )})$ and $\mathbf{k}_{p}=(k_{1},\ldots ,k_{p})$, $\mathbf{k}%
_{q}^{(\pm )}=(k_{p+2}^{(\pm )},\ldots ,k_{D}^{(\pm )})$. Now, the energy
corresponding to the modes reads%
\begin{equation}
\omega ^{(\pm )}=\sqrt{\mathbf{k}_{p}^{2}+k_{p+1}^{2}+(\mathbf{k}_{q}^{(\pm
)}\mp e\mathbf{A}_{q})^{2}+m^{2}}.  \label{ompm}
\end{equation}%
From the periodicity conditions (\ref{BC}), for the eigenvalues of the
momentum components along the compact dimensions we find%
\begin{equation}
k_{l}^{(\pm )}=2\pi (n_{l}\pm \alpha _{l})/L_{l},\;l=p+2,\ldots ,D,
\label{klpm}
\end{equation}%
with $n_{l}=0,\pm 1,\pm 2,\ldots $. For the components along the
uncompactified dimensions one has $-\infty <k_{l}<\infty $, $l=1,\ldots ,p$.

The relations between the coefficients in Eq. (\ref{phixi1}) is found from
the boundary condition (\ref{BagCond}) on the plate at $z=0$:
\begin{eqnarray}
\varphi _{+}^{(+)} &=&\frac{k_{p+1}\sigma _{p+1}\boldsymbol{\sigma }%
_{\parallel }^{+}\cdot (\mathbf{k}_{\parallel }^{(+)}-e\mathbf{A})-m(\omega
^{(+)}+m)-k_{p+1}^{2}}{(m-ik_{p+1})\left( \omega ^{(+)}+m\right) }\varphi
_{-}^{(+)},  \notag \\
\varphi _{-}^{(-)} &=&\frac{k_{p+1}\sigma _{p+1}^{+}\boldsymbol{\sigma }%
_{\parallel }\cdot (\mathbf{k}_{\parallel }^{(-)}+e\mathbf{A)-}m(\omega
^{(-)}+m)-k_{p+1}^{2}}{(m+ik_{p+1})(\omega ^{(-)}+m)}\varphi _{+}^{(-)},
\label{RelComp}
\end{eqnarray}%
where $\boldsymbol{\sigma }_{\parallel }=(\sigma _{1},\ldots ,\sigma
_{p},\sigma _{p+2},\ldots ,\sigma _{D})$. As we have already extracted the
coefficients $A_{\beta }^{(\pm )}$ in Eq. (\ref{modes}), we can impose the
normalization conditions $\varphi _{-}^{(+)+}\varphi _{-}^{(+)}=\varphi
_{+}^{(-)+}\varphi _{+}^{(-)}=1$. Now, from Eq. (\ref{RelComp}) it can be
seen that one has also the relations $\varphi _{+}^{(+)+}\varphi
_{+}^{(+)}=\varphi _{-}^{(-)+}\varphi _{-}^{(-)}=1$. As a set of independent
spinors with this normalization we take $\varphi _{-}^{(+)}=w^{(\sigma )}$
and $\varphi _{+}^{(-)}=w^{(\sigma )\prime }$. Here $w^{(\sigma )}$, $\sigma
=1,\ldots ,N_{D}/2$, are one-column matrices having $N_{D}/2$ rows with the
elements $w_{l}^{(\sigma )}=\delta _{l\sigma }$, and $w^{(\sigma )\prime
}=iw^{(\sigma )}$.

It remains for us to impose on the mode functions the boundary condition (\ref%
{BagCond}) on the plate at $z=a$. From this condition we find the following
equation for the eigenvalues of the component of the momentum normal to the
plates:
\begin{equation}
ma\sin (k_{p+1}a)/(k_{p+1}a)+\cos (k_{p+1}a)=0.  \label{NormMod}
\end{equation}%
We denote the positive solutions of this equation, arranged in the
ascending order, by $\lambda _{n}=k_{p+1}a$, $n=1,2,\ldots $. Note
that, for a massless
field, one has $\lambda _{n}=\pi (n-1/2)$. Hence, the mode functions (\ref%
{modes}) are specified by the set $\beta =(\mathbf{k}_{p},\mathbf{k}%
_{q}^{(\pm )},n,\sigma )$.

The coefficients $A_{\beta }^{(\pm )}$ in Eq. (\ref{modes}) are determined
from the normalization condition with the integration over the region
between the plates. They are given by the expression%
\begin{equation}
A_{\beta }^{(\pm )2}=\frac{\omega ^{(\pm )}+m}{4(2\pi )^{p}\omega ^{(\pm
)}aV_{q}}\left[ 1-\frac{\sin (2\lambda _{n})}{2\lambda _{n}}\right] ^{-1},
\label{Abeta}
\end{equation}%
with $V_{q}=L_{p+2}\cdots L_{D}$ being the volume of the compact subspace.

\section{Fermionic current}

\label{sec:FC}

Having the complete set of mode functions, we can evaluate the VEV
of the fermionic current by using the mode sum formula
(\ref{VEVj1}). First of all, we consider the $\mu =0$ component
which corresponds to the vacuum charge density.

\subsection{Charge density}

By taking into account Eq. (\ref{modes}) for the positive- and
negative-energy mode functions, for the charge density one finds the
expression below:%
\begin{eqnarray}
\langle j^{0}\rangle  &=&\frac{N_{D}}{4aV_{q}}\sum_{\mathbf{n}_{q}\in
\mathbf{Z}^{q}}\int \frac{d\mathbf{k}_{p}}{(2\pi )^{p}}\sum_{n=1}^{\infty }%
\left[ 1-\frac{\sin (2\lambda _{n})}{2\lambda _{n}}\right] ^{-1}  \notag \\
&&\times \left[ 2-ma\left( \frac{e^{2i\lambda _{n}z/a}}{ma-i\lambda _{n}}+%
\frac{e^{-2i\lambda _{n}z/a}}{ma+i\lambda _{n}}\right) \right] .  \label{j0}
\end{eqnarray}%
where $\mathbf{n}_{q}=(n_{p+2},\ldots ,n_{D})$, $-\infty
<n_{l}<+\infty $. This expression is not convenient for the direct
evaluation of the charge density. This is related to the fact that
the eigenvalues $\lambda _{n}$ are given implicitly and the terms
with large values of $n$ are highly oscillatory. These two
problems are solved by using the Abel-Plana-type summation formula
\cite{Rome02,Saha08Rev}
\begin{equation}
\sum_{n=1}^{\infty }\frac{\pi f(\lambda _{n})}{1-\sin (2\lambda
_{n})/(2\lambda _{n})}=-\frac{\pi maf(0)}{2(ma+1)}+\int_{0}^{\infty
}dx\,f(x)-i\int_{0}^{\infty }dx\frac{f(ix)-f(-ix)}{\frac{x+ma}{x-ma}e^{2x}+1}%
.  \label{Abel-Plan}
\end{equation}%
The corresponding function $f(x)$ in Eq. (\ref{j0}) is an even function and,
hence, the last integral in Eq. (\ref{Abel-Plan}) vanishes. In addition, one
has $f(0)=0$. Further, for the part in the first integral in the right-hand
side of Eq. (\ref{Abel-Plan}) we have%
\begin{equation}
\int_{0}^{\infty }dx\left( \frac{e^{-2ixz_{p+1}}}{m+ix}+\frac{e^{2ixz_{p+1}}%
}{m-ix}\right) =0.  \label{IntRel1}
\end{equation}%
Hence, after applying Eq. (\ref{Abel-Plan}) we get:%
\begin{equation}
\langle j^{0}\rangle =\frac{N_{D}}{2aV_{q}}\int \frac{d\mathbf{k}_{p+1}}{%
(2\pi )^{p+1}}\sum_{\mathbf{n}_{q}\in \mathbf{Z}^{q}}1\,.  \label{j01}
\end{equation}%
The latter coincides with the charge density in the spacetime with topology $%
R^{p+1}\times (S^{1})^{q}$, in the absence of boundaries. The
renormalized value of the latter vanishes \cite{Bell10}. Hence, we
conclude that the presence of boundaries with MIT bag boundary
conditions does not induce any vacuum charge density.

\subsection{Current density}

Next we consider the spatial components for $\langle j^{\mu
}\rangle $, $\mu =1,\ldots ,D$. Firstly, let us consider the
component of the current along the direction normal to the plates.
From Eqs. (\ref{VEVj1}) and (\ref{modes})
the following expression is obtained%
\begin{equation}
\langle j^{p+1}\rangle =\frac{i}{2}\sum_{j=+,-}\sum_{\beta }\frac{A_{\beta
}^{(j)2}}{\omega ^{(j)}+m}[\varphi ^{(j)+}\partial _{p+1}\varphi ^{(j)}-(%
\mathbf{\partial }_{p+1}\varphi ^{(j)+})\varphi ^{(j)}].  \label{jnorm}
\end{equation}%
By making use of Eq. (\ref{phixi1}), it can be seen that%
\begin{equation}
\varphi ^{(j)+}\partial _{p+1}\varphi ^{(j)}-(\mathbf{\partial }%
_{p+1}\varphi ^{(j)+})\varphi ^{(j)}=2ik_{p+1}(\varphi _{+}^{(j)+}\varphi
_{+}^{(j)}-\varphi _{-}^{(j)+}\varphi _{-}^{(j)})=0.  \label{RelNorm}
\end{equation}%
Hence, we conclude that the VEV of the normal component of the fermionic
current vanishes.

Now we turn to the components of the current along the directions parallel
to the plates. From Eqs. (\ref{VEVj1}) and (\ref{modes}) we get
\begin{equation*}
\langle j^{l}\rangle =\sum_{j=+,-}\sum_{\beta }\frac{A_{\beta }^{(j)2}}{%
\omega ^{(j)}+m}[-jk_{l}^{(j)}\varphi ^{(j)+}\varphi ^{(j)}+\frac{i}{2}%
\partial _{p+1}(\varphi ^{(j)+}\sigma _{l}^{(j^{\prime })}\sigma
_{p+1}^{(j)}\varphi ^{(j)})],
\end{equation*}%
where $l=1,\ldots ,p,p+2,\ldots D$, $j^{\prime }=+$ for $j=-$ and $j^{\prime
}=-$ for $j=+$. We have also used the notations $\sigma _{\mu }^{(-)}=\sigma
_{\mu }$ and $\sigma _{\mu }^{(+)}=\sigma _{\mu }^{+}$. By using Eqs. (\ref%
{phixi1}) and (\ref{RelComp}), after lengthy calculations, the
following representation is obtained for the corresponding mode
sum in the region
between the plates:%
\begin{eqnarray}
\langle j^{l}\rangle  &=&\frac{N_{D}}{4aV_{q}}\sum_{\mathbf{n}_{q}\in
\mathbf{Z}^{q}}\int \frac{d\mathbf{k}_{p}}{(2\pi )^{p}}\sum_{n=1}^{\infty }%
\left[ 1-\frac{\sin (2\lambda _{n})}{2\lambda _{n}}\right] ^{-1}  \notag \\
&&\times \frac{k_{l}}{\omega }\left[ 2-ma\left( \frac{e^{2i\lambda _{n}z/a}}{%
ma-i\lambda _{n}}+\frac{e^{-2i\lambda _{n}z/a}}{ma+i\lambda _{n}}\right) %
\right] ,  \label{jpar}
\end{eqnarray}%
where%
\begin{equation}
k_{l}=2\pi (n_{l}+\tilde{\alpha}_{l})/L_{l}\text{ for }l=p+2,\ldots ,D,
\label{klcomp}
\end{equation}%
and
\begin{equation}
\omega =\sqrt{\lambda _{n}^{2}/a^{2}+\mathbf{k}_{p}^{2}+\mathbf{k}%
_{q}^{2}+m^{2}},  \label{omeg}
\end{equation}%
with $\mathbf{k}_{q}=(k_{p+2},\ldots ,k_{D})$. In Eq. (\ref{klcomp}) we have
defined%
\begin{equation}
\tilde{\alpha}_{l}=eA_{l}L_{l}/(2\pi )-\alpha _{l}.  \label{alftilde}
\end{equation}%
Hence, the VEV of the fermionic current depends on the phases in the
periodicity conditions and on the components of the vector potential in the
combination (\ref{alftilde}). From Eq. (\ref{jpar}) it follows that the VEV\
is a periodic odd function of this parameter with the period equal to 1. In
particular, the fermionic current is a periodic function of $A_{l}L_{l}$
with the period of the flux quantum $\Phi _{0}=2\pi /|e|$ ($\Phi _{0}=2\pi
\hbar c/|e|$ in standard units).

From Eq. (\ref{jpar}) it directly follows that the components of the current
density along the uncompactified dimensions vanish:%
\begin{equation}
\langle j^{l}\rangle =0\text{ for}\;l=1,\ldots ,p.  \label{juncomp}
\end{equation}%
We also can see that the fermionic current along the $l$-th compact
dimension vanishes for the special cases $\tilde{\alpha}_{l}=0$ and $\tilde{%
\alpha}_{l}=1/2$. In particular, this is the case for untwisted and twisted
fermion fields in the absence of the gauge field.

By the reasons given after Eq. (\ref{j0}), formula (\ref{jpar}) is
not convenient for the investigation of the properties of the
current. In order to obtain a more workable representation, for
the summation of the series over $n$ we use Eq. (\ref{Abel-Plan}).
As a result, the VEV of the current
is decomposed as:%
\begin{eqnarray}
\langle j^{l}\rangle  &=&\langle j^{l}\rangle ^{(0)}+\langle j^{l}\rangle
^{(1)}-\frac{N_{D}}{2\pi V_{q}}\sum_{\mathbf{n}_{q}\in \mathbf{Z}%
^{q}}k_{l}\int \frac{d\mathbf{k}_{p}}{(2\pi )^{p}}\int_{\sqrt{k_{\parallel
}^{2}+m^{2}}}^{\infty }dx  \notag \\
&&\times \left. \frac{(x^{2}-k_{\parallel }^{2}-m^{2})^{-1/2}}{\frac{x+m}{x-m%
}e^{2ax}+1}\left[ 2-m\left( \frac{e^{2zx}}{m-x}+\frac{e^{-2zx}}{m+x}\right) %
\right] \right\} ,  \label{jpar1}
\end{eqnarray}%
where $k_{\parallel }^{2}=\mathbf{k}_{p}^{2}+\mathbf{k}_{q}^{2}$. Note that
the first two terms in the right-hand side of Eq. (\ref{jpar1}) come from
the first integral in the right-hand side of Eq. (\ref{Abel-Plan}). In Eq. (\ref{jpar1}%
),
\begin{equation}
\langle j^{l}\rangle ^{(0)}=\frac{N_{D}}{2V_{q}}\sum_{\mathbf{n}_{q}\in
\mathbf{Z}^{q}}\int \frac{d\mathbf{k}_{p+1}}{(2\pi )^{p+1}}\frac{k_{l}}{%
\sqrt{\mathbf{k}_{p+1}^{2}+\mathbf{k}_{q}^{2}+m^{2}}},  \label{jt}
\end{equation}%
is the corresponding fermionic current in the boundary-free spacetime with
topology $R^{p+1}\times (S^{1})^{q}$ and the part%
\begin{eqnarray}
\langle j^{l}\rangle ^{(1)} &=&-\frac{N_{D}m}{4\pi V_{q}}\int \frac{d\mathbf{%
k}_{p}}{(2\pi )^{p}}\sum_{\mathbf{n}_{q}\in \mathbf{Z}^{q}}k_{l}\int_{0}^{%
\infty }dx  \notag \\
&&\times \frac{e^{2ixz}/(m-ix)+e^{-2ixz}/(m+ix)}{\sqrt{x^{2}+\mathbf{k}%
_{p}^{2}+\mathbf{k}_{q}^{2}+m^{2}}},  \label{j11}
\end{eqnarray}%
is induced by the presence of a single plate located at $z=0$,
when the second plate is absent. The latter can be seen by the
direct evaluation of the VEV for a single plate, or by taking the
limit $a\rightarrow \infty $ in Eq. (\ref{jpar1}). Hence, the last
term in Eq. (\ref{jpar1}) can be interpreted as being induced by
the presence of the second plate at $z=a$.

The boundary induced parts in Eq. (\ref{jpar1}) are finite (see below) and
the renormalization is necessary for the boundary-free part only. The latter
has been investigated in Ref. \cite{Bell10}. The corresponding renormalized
VEV is given by the expression
\begin{eqnarray}
\langle j^{l}\rangle ^{(0)} &=&\frac{2N_{D}m^{D+1}L_{l}}{(2\pi )^{(D+1)/2}}%
\sum_{n_{l}=1}^{\infty }n_{l}\sin (2\pi n_{l}\tilde{\alpha}_{l})  \notag \\
&&\times \sum_{\mathbf{n}_{q-1}\in \mathbf{Z}^{q-1}}\cos (2\pi \mathbf{n}%
_{q-1}\cdot \boldsymbol{\alpha }_{q-1})\frac{K_{(D+1)/2}(mg(\mathbf{L}_{q},%
\mathbf{n}_{q}))}{(mg(\mathbf{L}_{q},\mathbf{n}_{q}))^{(D+1)/2}},
\label{jr3}
\end{eqnarray}%
where $\mathbf{n}_{q-1}=(n_{p+2},\ldots ,n_{l-1},n_{l+1},\ldots ,n_{D})$, $%
\boldsymbol{\alpha }_{q-1}=(\tilde{\alpha}_{p+1},\ldots ,\tilde{\alpha}%
_{l-1},\tilde{\alpha}_{l+1},\ldots \tilde{\alpha}_{D})$, $K_{\nu
}(x)$ is
the Macdonald function (i.e. the modified Bessel function of the second kind) and%
\begin{equation}
g(\mathbf{L}_{q},\mathbf{n}_{q})=\Big(\sum_{i=p+2}^{D}L_{i}^{2}n_{i}^{2}\Big)%
^{1/2}.  \label{glqng}
\end{equation}%
The reason for the sign difference of $\alpha _{l}$ in the expression of $%
\tilde{\alpha}_{l}$ in Eq. (\ref{alftilde}) and in the
corresponding expression of Ref. \cite{Bell10} is that, in the
latter reference, for the evaluation of the VEV the
negative-energy modes have been used, with the eigenvalues
$k_{l}^{(+)}$ (see Eq. (\ref{klpm})) instead of $k_{l}^{(-)}$.
This means that, in fact, the formulas given in Ref. \cite{Bell10}
are for
the periodicity conditions (\ref{BC}) with $\alpha _{l}$ replaced by $%
-\alpha _{l}$. The formula (\ref{jr3}) is obtained by making use of the zeta
function technique. An alternative representation, derived by using the
Abel-Plana-type summation formula, is given in Rev. \cite{Bell10}. In the
discussion below we will be mainly concerned with the effects induced by the
plates.

\subsubsection{Fermionic current induced by a single plate}

First of all, we discuss the part in the fermionic current induced
by a single plate. As it is seen from Eq. (\ref{j11}), for a
massless field the latter vanishes. In the case of a massive
field, for the further transformation of Eq. (\ref{j11}), we
rotate the integration contour in the complex plane $x$ by the
angle $\pi /2$ for the term with $e^{2ixz}$ and by the angle $-\pi
/2$
for the term with $e^{2ixz}$. This gives the result:%
\begin{equation}
\langle j^{l}\rangle ^{(1)}=-\frac{N_{D}m}{(2\pi )^{p+1}V_{q}}\sum_{\mathbf{n%
}_{q}\in \mathbf{Z}^{q}}k_{l}\int d\mathbf{k}_{p}\int_{\sqrt{\mathbf{k}%
_{p}^{2}+m_{\mathbf{n}_{q}}^{2}}}^{\infty }dx\frac{e^{-2xz}/(m+x)}{\sqrt{%
x^{2}-\mathbf{k}_{p}^{2}-m_{\mathbf{n}_{q}}^{2}}},  \label{j11b}
\end{equation}%
where%
\begin{equation}
m_{\mathbf{n}_{q}}=\sqrt{\mathbf{k}_{q}^{2}+m^{2}}.  \label{mnq}
\end{equation}%
This expression is further simplified by using the relation%
\begin{equation}
\int d\mathbf{k}_{p}\int_{\sqrt{\mathbf{k}_{p}^{2}+m_{\mathbf{n}_{q}}^{2}}%
}^{\infty }\frac{f(x)dx}{\sqrt{x^{2}-\mathbf{k}_{p}^{2}-m_{\mathbf{n}%
_{q}}^{2}}}=\frac{\pi ^{(p+1)/2}}{\Gamma ((p+1)/2)}\int_{b}^{\infty
}dy\,(y^{2}-m_{\mathbf{n}_{q}}^{2})^{(p-1)/2}f(y).  \label{IntRel2}
\end{equation}%
The final formula can be written in a form valid for both plates
and for
both sides of the plate:%
\begin{equation}
\langle j^{l}\rangle _{j}^{(1)}=-\frac{A_{p}}{V_{q}}N_{D}m\sum_{\mathbf{n}%
_{q}\in \mathbf{Z}^{q}}k_{l}\int_{m_{\mathbf{n}_{q}}}^{\infty }dx\,(x^{2}-m_{%
\mathbf{n}_{q}}^{2})^{(p-1)/2}\frac{e^{-2x|z-a_{j}|}}{m+x},  \label{j12}
\end{equation}%
with the notation%
\begin{equation}
A_{p}=\frac{(4\pi )^{-(p+1)/2}}{\Gamma ((p+1)/2)}.  \label{Ap}
\end{equation}%
In Eq. (\ref{j12}), $j=1,2$, $a_{1}=0$, $a_{2}=a$, and $\langle j^{l}\rangle
_{j}^{(1)}$ is the VEV induced by a single plate placed at $z=a_{j}$, so $%
\langle j^{l}\rangle _{1}^{(1)}=\langle j^{l}\rangle ^{(1)}$. Note that the
integral in Eq. (\ref{j12}) is a monotonically decreasing positive function
of $m_{\mathbf{n}_{q}}$. An alternative expression for the plate induced
part in the VEV of the fermionic current is derived in Appendix \ref%
{sec:Appendix}, by using the summation formula (\ref{SumForm}).

If the length of the $r$-th compactified dimension, $L_{r}$, $r\neq l$, is
large compared to the other length scales, the dominant contribution to the
sum over $n_{r}$ in Eq.~(\ref{j12}) comes from large values of $n_{r}$. In
this case, to the leading order, we can replace the corresponding series by
an integral, using the relation%
\begin{equation}
\sum_{n_{r}=-\infty }^{+\infty }f(2\pi |n_{r}+\tilde{\alpha}%
_{r}|/L_{r})\rightarrow \frac{L_{r}}{\pi }\int_{0}^{\infty }dy\,f(y).
\label{SumInt}
\end{equation}%
The double integral is reduced to the single one with the help of the formula%
\begin{equation}
\int_{0}^{\infty }dy\,\int_{\sqrt{y^{2}+b^{2}}}^{\infty
}dx\,(x^{2}-y^{2}-b^{2})^{(p-1)/2}f(x)=\frac{\sqrt{\pi }\Gamma ((1+p)/2)}{%
2\Gamma (1+p/2)}\int_{b}^{\infty }du(u^{2}-b^{2})^{p/2}\,f(u),
\label{IntRel3}
\end{equation}%
and from Eq. (\ref{j12}) the corresponding expression is obtained
for the topology $R^{p+2}\times (S^{1})^{q-1}$. Note that this way
of calculating does not work for the case $r=l$, as the leading
term obtained by the replacement (\ref{SumInt}) vanishes. The
asymptotic expression for large values of $L_{l}$ will be given
below, by using the alternative representation of the boundary
induced part in the VEV given in Appendix.

In the opposite limit, i.e. when $L_{r}$ is small, the behavior of
the current density depends crucially whether the parameter
$\tilde{\alpha}_{r}$ is zero
or not. For $\tilde{\alpha}_{r}=0$ the dominant contribution in Eq. (\ref%
{j12}) comes from the mode with $n_{r}=0$, and to the leading order one has%
\begin{equation}
\langle j^{l}\rangle _{j}^{(1)}\approx N_{D}(N_{D-1}L_{r})^{-1}\langle
j^{l}\rangle _{j,R^{p+1}\times (S^{1})^{q-1}}^{(1)},  \label{j1as}
\end{equation}%
where $\langle j^{l}\rangle _{j,R^{p+1}\times (S^{1})^{q-1}}^{(1)}$ is the
corresponding quantity in $(D-1)$-dimensional space with topology $%
R^{p+1}\times (S^{1})^{q-1}$ and with the lengths of the compact dimensions $%
L_{p+2},\ldots ,L_{r-1},L_{r+1},\ldots ,L_{D}$. For $\tilde{\alpha}_{r}\neq 0
$, the dominant contribution comes from the region near the lower limit of
the integration in Eq. (\ref{j12}) and we get:%
\begin{equation}
\langle j^{l}\rangle _{j}^{(1)}\approx -\frac{N_{D}m}{V_{q}}\frac{%
|z-a_{j}|^{-(p+1)/2}}{2(4\pi )^{(p+1)/2}}\sum_{\mathbf{n}_{q}\in \mathbf{Z}%
^{q}}k_{l}m_{n_{\mathbf{q}}}^{(p-3)/2}e^{-2|z-a_{j}|m_{n_{\mathbf{q}}}}.
\label{SmallLr}
\end{equation}%
As before, the contribution of the term with $n_{r}=0$ dominates
and the VEV
is exponentially suppressed by the factor $e^{-4\pi \tilde{\alpha}%
_{r}|z-a_{j}|/L_{r}}$. Note that this asymptotic form is valid for
$r=l$ as well.

For the investigation of the boundary induced part near the plate
and for large values of $L_{l}$, it is more convenient to use Eq.
(\ref{j15}). From this expression it is seen that the boundary
induced part is finite on the plate and the corresponding value is
obtained by the direct substitution $z=0 $. This contrasts with
the case of the fermionic condensate and the VEV of the
energy-momentum tensor (see Ref. \cite{Eliz11}) which are
divergent on the boundary.

In the limit of large values of $L_{l}$ the dominant contribution in Eq. (%
\ref{j15}) comes from the region near the lower limit of the integral over $x
$. Assuming $|\tilde{\alpha}_{l}|<1/2$, we can see that the contribution of
the term with $n_{i}=0$, $i=p+2,\ldots ,l-1,l+1,\ldots ,D$, dominates in the
sum over $\mathbf{n}_{q-1}$. To the leading order we get%
\begin{equation}
\langle j^{l}\rangle _{j}^{(1)}\approx -\frac{N_{D}L_{l}}{V_{q}}\frac{\sin
(2\pi \tilde{\alpha}_{l})}{(2\pi L_{l})^{p/2+1}}m_{l}^{p/2+1}e^{-m_{l}L_{l}},
\label{j1LargeLl}
\end{equation}%
with the notation%
\begin{equation}
m_{l}^{2}=\sum_{i=p+2,i\neq l}^{D}(2\pi \tilde{\alpha}_{i}/L_{i})^{2}+m^{2}.
\label{ml2}
\end{equation}%
Hence, for large values of the length of the compact dimension we
have an exponential suppression. In this limit the total VEV\ is
dominated by the boundary-free part $\langle j^{l}\rangle ^{(0)}$.
As we see, in both limits of small and large values of $L_{l}$ the
boundary induced part in the VEV of the fermionic current goes to
zero.

\begin{figure}[tbp]
\begin{center}
\begin{tabular}{cc}
\epsfig{figure=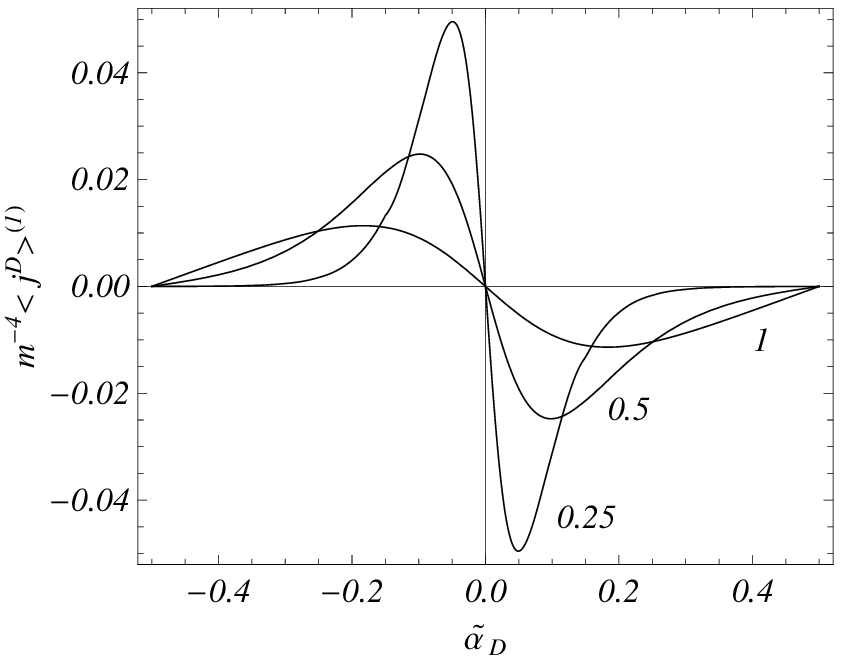,width=7.cm,height=6.cm} & \quad %
\epsfig{figure=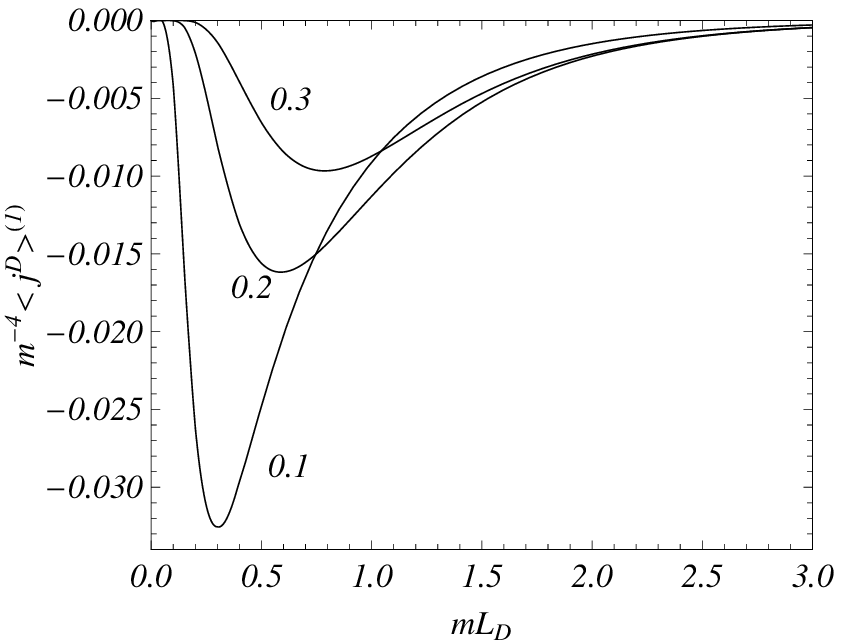,width=7.cm,height=6.cm}%
\end{tabular}%
\end{center}
\caption{VEV of the current density induced by a single plate in
the Kaluza-Klein model with spatial topology $R^{3}\times S^{1}$,
as a function of the parameter $\tilde{\protect\alpha}_{D}$ (left
plot) and of the parameter $mL_{D}$ (right plot), for a fixed
distance from the plate
corresponding to $mz=0.5$. The numbers near the curves are the values of $%
mL_{D}$ (left plot) and $\tilde{\protect\alpha}_{D}$ (right
plot).} \label{fig1}
\end{figure}

In the numerical examples below we consider the simplest Kaluza-Klein model
with a single compact dimension with $p=D-2$, $q=1$, and $k_{\mathbf{n}%
_{q}}=2\pi |n_{D}+\alpha _{D}|/L_{D}$. In the left panel of Fig.
\ref{fig1}, the current density induced by a single plate at $z=0$
is plotted versus the parameter $\tilde{\alpha}_{D}$, for a fixed
value $mz=0.5$. The numbers near the curves are the corresponding
values of $mL_{D}$ (the length of the compact space in units of
the Compton wavelength of the fermionic particle). As we already
mentioned before, the current density is a periodic function of
the parameter $\tilde{\alpha}_{D}$ with the period equal to 1.
As it is seen from the graphs, the absolute value of the parameter $\tilde{%
\alpha}_{D}$, for which the current is maximum, increases with
increasing the length of the compact dimension. The value of the
current at the maximum decreases with increasing the length. The
right panel of Fig. \ref{fig1} presents the dependence of the
fermionic current on the length of the compact dimension for
separate values of the parameter $\tilde{\alpha}_{D}$ (numbers
near the curves) and for $mz=0.5$. As we already mentioned, the
boundary induced part vanishes in both limits of small and large
values of the compact dimension length.

\subsubsection{Geometry of two plates}

By using Eq. (\ref{IntRel2}), the total VEV in the region between two plates
is presented in the form%
\begin{eqnarray}
\langle j^{l}\rangle  &=&\langle j^{l}\rangle ^{(0)}+\langle j^{l}\rangle
^{(1)}-N_{D}\frac{A_{p}}{V_{q}}\sum_{\mathbf{n}_{q}\in \mathbf{Z}%
^{q}}k_{l}\int_{m_{\mathbf{n}_{q}}}^{\infty }dx  \notag \\
&&\times \frac{(x^{2}-m_{\mathbf{n}_{q}}^{2})^{(p-1)/2}}{\frac{x+m}{x-m}%
e^{2ax}+1}\left[ 2-m\left( \frac{e^{2zx}}{m-x}+\frac{e^{-2zx}}{m+x}\right) %
\right] ,  \label{jpar2}
\end{eqnarray}%
where $m_{\mathbf{n}_{q}}$ is defined by Eq. (\ref{mnq}). Combining with Eq.
(\ref{j12}) for the geometry of a single plate, we find%
\begin{eqnarray}
\langle j^{l}\rangle  &=&\langle j^{l}\rangle ^{(0)}-2N_{D}\frac{A_{p}}{V_{q}%
}\sum_{\mathbf{n}_{q}\in \mathbf{Z}^{q}}k_{l}\int_{m_{\mathbf{n}%
_{q}}}^{\infty }dx\,  \notag \\
&&\times \frac{(x^{2}-m_{\mathbf{n}_{q}}^{2})^{(p-1)/2}}{\frac{x+m}{x-m}%
e^{2ax}+1}\left\{ 1+\frac{me^{ax}}{x-m}\cosh \left[ (a-2z)x\right] \right\} .
\label{jpar3}
\end{eqnarray}%
As we could expect, this expression is symmetric with respect to the plane $%
z=a/2$. The integral in Eq. (\ref{jpar3}) is a monotonically decreasing
positive function of $m_{\mathbf{n}_{q}}$.

The formula (\ref{jpar3})\ is further simplified for a massless field:%
\begin{equation}
\langle j^{l}\rangle =\langle j^{l}\rangle ^{(0)}-2N_{D}\frac{A_{p}}{V_{q}}%
\sum_{\mathbf{n}_{q}\in \mathbf{Z}^{q}}k_{l}\int_{k_{\mathbf{n}%
_{q}}}^{\infty }dx\frac{(x^{2}-k_{\mathbf{n}_{q}}^{2})^{(p-1)/2}}{e^{2ax}+1},
\label{jparm0}
\end{equation}%
where%
\begin{equation}
k_{\mathbf{n}_{q}}^{2}=\sum_{i=p+2}^{D}[2\pi (n_{i}+\tilde{\alpha}%
_{i})/L_{i}]^{2}.  \label{knq}
\end{equation}%
Note that in this case the distribution of the fermionic current in the
region between the plates is uniform. An alternative expression is obtained
by making use of the expansion $(e^{y}+1)^{-1}=-\sum_{n=1}^{\infty
}(-1)^{n}e^{-ny}$. Then the integral is expressed in terms of the Macdonald
function and one gets%
\begin{equation}
\langle j^{l}\rangle =\langle j^{l}\rangle ^{(0)}+\frac{4N_{D}a^{-p/2}}{%
(4\pi )^{p/2+1}V_{q}}\sum_{\mathbf{n}_{q}\in \mathbf{Z}^{q}}k_{l}k_{\mathbf{n%
}_{q}}^{p/2}\sum_{n=1}^{\infty }\frac{(-1)^{n}}{n^{p/2}}K_{p/2}(2nak_{%
\mathbf{n}_{q}}).  \label{jparm0B}
\end{equation}%
At large separations between the plates, compared with the lengths of the
compact directions, the dominant contribution to Eq. (\ref{jparm0B}) comes
from the terms with $n_{i}=0$ for $\left\vert \tilde{\alpha}_{i}\right\vert
<1/2$ and from the terms $n_{i}=0,\pm 1$ for $\tilde{\alpha}_{i}=\mp 1/2$.
For $\left\vert \tilde{\alpha}_{l}\right\vert \leqslant 1/2$, to the leading
order one finds:%
\begin{equation}
\langle j^{l}\rangle \approx \langle j^{l}\rangle ^{(0)}-\frac{%
2^{N_{1/2}}N_{D}\tilde{\alpha}_{l}\beta _{0}^{(p-1)/2}}{%
(2a)^{(p+1)/2}V_{q}L_{l}}e^{-4\pi a\beta _{0}},  \label{jlLargeSep}
\end{equation}%
where $\beta _{0}^{2}=\sum_{i=p+2}^{D}\left( \tilde{\alpha}_{i}/L_{i}\right)
^{2}$ and $N_{1/2}$ is the number of compact dimensions for which $\tilde{%
\alpha}_{i}=\pm 1/2$, $i\neq l$. As it is seen, in this limit the boundary
induced part in the VEV is exponentially small.

Extracting the parts corresponding to the single plates, the fermionic
current can also be presented in the form%
\begin{equation}
\langle j^{l}\rangle =\langle j^{l}\rangle ^{(0)}+\sum_{j=1,2}\langle
j^{l}\rangle _{j}^{(1)}+\Delta \langle j^{l}\rangle ,  \label{jinterf}
\end{equation}%
where the interference term is given by the expression%
\begin{eqnarray}
\Delta \langle j^{l}\rangle  &=&-\frac{A_{p}N_{D}}{V_{q}}\sum_{\mathbf{n}%
_{q}\in \mathbf{Z}^{q}}k_{l}\int_{m_{n_{\mathbf{q}}}}^{\infty }dx\frac{%
(x^{2}-m_{n_{\mathbf{q}}}^{2})^{(p-1)/2}}{\frac{x+m}{x-m}e^{2ax}+1}  \notag
\\
&&\times \left[ 2-\frac{m}{x+m}\left( e^{-2zx}+e^{2x(z-a)}\right) \right] .
\label{Deltj}
\end{eqnarray}%
For a massless field the interference part coincides with the
boundary induced part. In the limit of large separation between
the plates, the dominant contribution comes from the region near
the lower limit of the integration in Eq. (\ref{Deltj}). To the
leading order we get
\begin{eqnarray}
\Delta \langle j^{l}\rangle  &\approx &-\frac{2^{N_{1/2}}\pi N_{D}\tilde{%
\alpha}_{l}m_{0}^{(p-1)/2}}{V_{q}(4\pi a)^{(p+1)/2}L_{l}}\frac{m_{0}-m}{%
m_{0}+m}e^{-2am_{0}}  \notag \\
&&\times \left[ 2-\frac{m}{m_{0}+m}\left(
e^{-2zm_{0}}+e^{2m_{0}(z-a)}\right) \right] ,  \label{jlLargSepm}
\end{eqnarray}%
where%
\begin{equation}
m_{0}^{2}=\sum_{i=p+2}^{D}\left( 2\pi \tilde{\alpha}_{i}/L_{i}\right)
^{2}+m^{2}.  \label{m0}
\end{equation}%
For a massless field this result is reduced to Eq.
(\ref{jlLargeSep}). Hence, in the limit under consideration, the
interference effects between the boundaries are exponentially
suppressed.

Up to now, we have considered the region between the plates, $0<z<a$. For
the regions $z<0$ and $z>a$, the VEV of the components of the fermionic
current along the compact dimension $z_{l}$ is given by the expressions:%
\begin{eqnarray}
\langle j^{l}\rangle  &=&\langle j^{l}\rangle ^{(0)}+\langle j^{l}\rangle
_{1}^{(1)},\;z<0,  \notag \\
\langle j^{l}\rangle  &=&\langle j^{l}\rangle ^{(0)}+\langle j^{l}\rangle
_{2}^{(1)},\;z>a,  \label{jlout}
\end{eqnarray}%
with the plate induced part given by Eq.~(\ref{j12}). In
particular, the boundary induced part vanishes in these regions,
for a massless field.

\begin{figure}[tbp]
\begin{center}
\begin{tabular}{cc}
\epsfig{figure=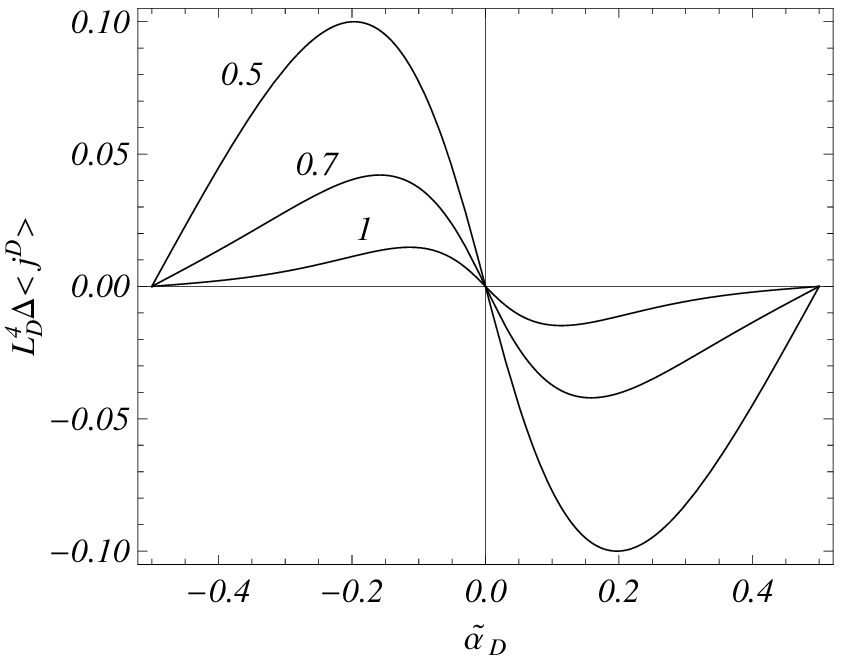,width=7.cm,height=6.cm} & \quad %
\epsfig{figure=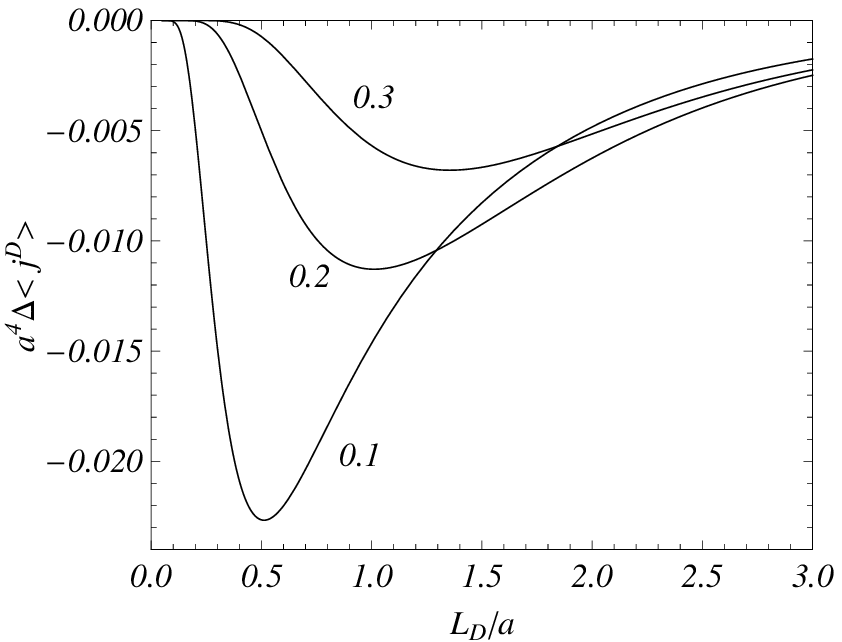,width=7.cm,height=6.cm}%
\end{tabular}%
\end{center}
\caption{Boundary induced part of the fermionic current versus $\tilde{%
\protect\alpha}_{D}$ (left plot) and $L_{D}/a$ (right plot) in the region
between two plates for the model with spatial topology $R^{3}\times S^{1}$
and for a massless field. The numbers near the curves are the values of $%
a/L_{D}$ (left plot) and of $\tilde{\protect\alpha}_{D}$ (right plot).}
\label{fig2}
\end{figure}

As in the case of a single plate, in the numerical example we
consider the model with spatial topology $R^{3}\times S^{1}$ and
with the length of the compact dimension $L_{D}$. In
Fig.~\ref{fig2}, the boundary induced part in the VEV of the
corresponding fermionic current is plotted for a massless
field in the region between two plates, as a function of the parameter $%
\tilde{\alpha}_{D}$ (left plot) and as a function of $L_{D}/a$
(right plot). The numbers near the curves correspond to the values
of the ratio $a/L_{D}$ for the left plot and to the values of
$\tilde{\alpha}_{D}$ for the right plot. Similar to the single
plate case, the boundary induced part vanishes in both limits of
small and large values of the compact dimension length.

\subsection{Fermionic current in conformally-flat spacetimes}

From quantum field theory in curved spacetime it is well-known
(see, for instance, Ref. \cite{Birr82}) that, in conformally-flat
spacetimes, the expectation values of bilinear field combinations
for conformally invariant fields are related to the corresponding
quantities in flat spacetime by a conformal transformation. By
taking into account that a massless fermionic field is conformally
invariant in all spatial dimensions, we can use the results given
above for the generation of the topological and boundary induced
parts in the VEV of the fermionic current for conformally-flat
spacetimes with spatial topology $R^{p+1}\times (S^{1})^{q}$ and
with the
line-element%
\begin{equation}
ds^{2}=\Omega ^{2}(z_{l})[dt^{2}-\sum_{i=1}^{D}(dz_{i})^{2}],
\label{ds2conf}
\end{equation}%
where, as before, $0\leqslant z_{l}\leqslant L_{l}$, $l=p+2,\ldots
,D$. The functions $\Omega (z_{l})$ are assumed to be periodic
along the compact dimensions. The Dirac matrices in this
spacetime, $\gamma _{(\Omega )}^{\mu } $, are related to the
matrices in flat spacetime by $\gamma _{(\Omega )}^{\mu
}=e_{l}^{\mu }\gamma ^{l}$, where $e_{l}^{\mu }=\Omega ^{-1}\delta
_{l}^{\mu }$ is the corresponding tetrad field. If $\psi _{(\Omega
)}$ is a massless fermion field in the spacetime with the
line-element (\ref{ds2conf}), then we have the following conformal
transformation: $\psi _{(\Omega )}=\Omega ^{-D/2}\psi $,
$\bar{\psi}_{(\Omega )}=\Omega ^{-D/2}\bar{\psi}$. By taking into
account that, for the normal to the boundaries, one has
$n_{(\Omega )\mu }=\Omega n_{\mu }$, we see that the bag boundary
condition $[1+i\gamma _{(\Omega )}^{\mu }n_{(\Omega )\mu }]\psi
_{(\Omega )}=0$ is invariant under the conformal transformation.
Assuming the presence of two boundaries at $z=0 $ and $z=a$
($z_{p+1}\equiv z$), with bag boundary conditions, for the
fermionic current in the spacetime (\ref{ds2conf}), we get%
\begin{equation}
\langle j^{\mu }\rangle _{(\Omega )}=\langle j^{\mu }\rangle _{(\Omega
),R^{D}}+\Omega ^{-D-1}\langle j^{\mu }\rangle .  \label{jCurv}
\end{equation}%
Here, the first term in the right-hand side is the corresponding VEV for a
spacetime with the line-element (\ref{ds2conf}) and with spatial topology $%
R^{D}$, and the second term is induced by the nontrivial topology
and by the boundaries. With this form, the renormalization is
needed for the first term only. For the special cases of de Sitter
and anti-de Sitter (AdS) spacetimes, one has $\Omega
^{2}=1/(Ht)^{2}$ and $\Omega ^{2}=1/(kz)^{2}$, respectively.

Let us consider the case of AdS bulk in detail. The boundaries in AdS
spacetime can play the role of branes in higher-dimensional generalizations
of Randall-Sundrum-type braneworlds with toroidally compactified internal
spaces. In these models the coordinate $z$ is compactified on an orbifold $%
S^{1}/Z_{2}$, and the branes are on the two fixed points. The
left/right boundaries correspond to the hidden/visible branes. By
taking into account that $z=0$ represents the AdS boundary
($z=\infty $ corresponds to the AdS horizon), we will assume that
the boundaries are located at $z=a_{1}\neq 0$ and $z=a_{2}$,
$a_{1}<a_{2}$. The physical coordinate in the direction
perpendicular to the branes corresponds to $y=k^{-1}\ln (kz)$,
with $k^{-1}$ being the AdS radius. The physical distance between
the branes, in terms of this coordinate, is given by
$y_{0}=k^{-1}\ln (a_{2}/a_{1})$. To discuss the physics from the
point of view of a $D$-dimensional observer residing on the
visible brane, we introduce rescaled coordinates $z_{l}^{\prime }$
on this brane as $z_{l}^{\prime }=z_{l}/(ka_{2})$. With these
coordinates the warp factor in the metric (\ref{ds2conf}) is equal
to one at the brane $z=a_{2}$, and they are physical coordinates
for an observer on this brane. The corresponding lengths of the
compact dimensions are given by $L_{l}^{\prime }=L_{l}/(ka_{2})$.
For the part in the VEV of the fermionic current induced by the
hidden brane and measured by an observer on the visible brane,
from
Eqs. (\ref{jparm0B}) and (\ref{jCurv}) one finds%
\begin{eqnarray}
\langle j^{\prime l}\rangle _{\mathrm{AdS}}^{\mathrm{(b)}} &=& 4N_{D}k^{p/2}\frac{%
(1-e^{-ky_{0}})^{-p/2}}{(4\pi )^{p/2+1}V_{q}^{\prime }}\sum_{%
\mathbf{n}_{q}\in \mathbf{Z}^{q}}k_{l}^{\prime }k_{\mathbf{n}_{q}}^{\prime
p/2}  \notag \\
&&\times \sum_{n=1}^{\infty }\frac{(-1)^{n}}{n^{p/2}}%
K_{p/2}(2n(1-e^{-ky_{0}})k_{\mathbf{n}_{q}}^{\prime }/k),  \label{jlbrane}
\end{eqnarray}%
where the expressions for the primed quantities in the right-hand side are
obtained from the corresponding expressions above by the replacement $%
L_{l}\rightarrow L_{l}^{\prime }$. When the left brane tends to
the AdS boundary, $a_{1}\rightarrow 0$, the fermionic current goes
to a finite limiting value.

\section{Fermionic current in finite-length carbon nanotubes}

\label{sec:Nano}

In this section, we apply general formulas given above, for the
investigation of the fermionic current in finite-length carbon
nanotubes. Because of their unique electronic properties and
potential applications in nanotechnology, carbon nanotubes have
attracted enormous attention (see, for instance, Ref.
\cite{Sait98}). A single-walled cylindrical nanotube can be
thought of, as a graphene sheet rolled in a cylindrical form, with
diameters ranging from 1 nm to 5 nm. Multi-walled carbon nanotubes
are made of coaxial nanotube cylinders. The diameter of the outer
tubes for multi-walled nanotubes can be of the order of 500 nm.
The electronic coupling between separate layers in these nanotubes
is weak, and the results given below can be applied for both
single-walled and multi-walled carbon nanotubes. The low-energy
excitations of the electronic subsystem in a graphene sheet are
well described by the
(2+1)-dimensional Dirac model containing a pair of spinors $\psi _{A}$ and $%
\psi _{B}$ (for a review see Ref. \cite{Cast09}). The latter
correspond to the two different triangular sublattices of the
honeycomb lattice of graphene. Recently, a number of predictions
of the Dirac model have been confirmed experimentally with a high
precision. For carbon nanotubes the
background space for the (2+1)-dimensional Dirac model has the topology $%
R^{1}\times S^{1}$. The field equation for the separate spinors reads%
\begin{equation}
(iv_{F}^{-1}\gamma ^{0}D_{0}+i\gamma ^{l}D_{l}-m)\psi _{J}=0,\   \label{DGra}
\end{equation}%
where $J=A,B$ and $v_{F}\approx 10^{8}$ cm/s is the Fermi velocity of
electrons. The mass (gap) term in Eq.~(\ref{DGra}) can be generated by a
number of mechanisms (see, for example, Ref. \cite{Gusy95}).

The electrical properties of carbon nanotubes crucially depend on diameter
and chirality. Depending on the chirality, the nanotubes are either metallic
or semiconducting. The chirality also determines the periodicity condition
along the compact dimension for the fields $\psi _{J}$. In metallic
nanotubes, the periodic boundary condition is realized with the value $%
\alpha _{l}=0$ for the phase in Eq. (\ref{BC}). For semiconducting
nanotubes, depending on the chiral vector, there are two classes
of inequivalent periodicity conditions, corresponding to $\alpha
_{l}=\pm 1/3$. These phases have opposite signs for the
sublattices $A$ and $B$.

Although carbon nanotubes are typically longer than a micrometer,
recently a number of techniques have been developed for the
synthesis of ultra-short carbon nanotubes with lengths ranging
between 20 and 80 nm (see, for instance, Ref. \cite{Ashc06}). This
type of nanotubes are especially useful in biomedical
applications. In the long-wavelength description of the electronic
subsystem, the Dirac fields $\psi _{J}$ live on the cylinder
surface and for finite-length nanotubes additional boundary
conditions should be imposed at the edges. As the electrons are
confined between two edges of the nanotube, it is natural to
impose the bag boundary conditions at the edges. As we mentioned
before, the latter results in zero fermion flux through the
boundaries. The effect of these boundary conditions on the
fermionic current can be investigated by using the expressions
from the previous section. For nanotubes the corresponding values
for the dimensions are as follows: $D=2$, $p=0$, $q=1$. Another
type of boundary induced effect, i.e. the Casimir effect between
two graphene sheets or for a
graphene sheet interacting with a metal, has been investigated in Refs. \cite%
{Bord06}, using either the hydrodynamic model or the Dirac model
for quasiparticles in graphene (for a comparison of the results
obtained by these two approaches see Ref.~\cite{Chur10}).

Before considering finite-length effects, we recall the result for
the fermionic current in an infinite-length nanotube. The
corresponding
expression is obtained from Eq. (\ref{jr3}) with $\alpha _{l}=\alpha $ and $%
L_{2}=L$. Summing the contributions from the two sublattices and
taking into account that, for them the phases $\alpha $ have
opposite signs, for the total fermionic current we find
\cite{Bell10}
\begin{equation}
\langle j^{2}\rangle _{\mathrm{(cn)}}^{(0)}=\frac{2v_{F}}{\pi L^{2}}%
\sum_{n=1}^{\infty }\cos (2\pi n\alpha )\sin (2\pi n\Phi /\Phi _{0})\frac{%
1+nLm}{n^{2}e^{nLm}},  \label{jrs}
\end{equation}%
where $\alpha =0,1/3$ for metallic and semiconducting nanotubes,
respectively. In Eq. (\ref{jrs}), we have expressed the component
of the vector potential along the compact dimension in terms of
the magnetic flux $\Phi $ passing
through the cross section of the nanotube: $\Phi =A_{2}L$. Note that in Eq. (%
\ref{jrs}) and in the formulas below, we give the fermionic
current for a given spin component. The total current is obtained
multiplying by the number of spin components, which is 2 for
graphene. From Eq. (\ref{jrs}) it follows that, in the absence of
the magnetic flux, the total fermionic current vanishes, due to
the cancellation of contributions from the two sublattices.

Next, we consider a semi-infinite nanotube which corresponds to
the single plate geometry discussed in the previous section.
Summing the contributions
from two different sublattices, from the general formula (\ref{j12}) one gets:%
\begin{equation}
\langle j^{2}\rangle _{\mathrm{(cn)}}^{(1)}=-\frac{mv_{F}}{\pi L}%
\sum_{j=+,-}\sum_{n=-\infty }^{+\infty
}k_{n}^{(j)}\int_{m_{n}^{(j)}}^{\infty }dx\,\frac{e^{-2xz}}{\sqrt{%
x^{2}-m_{n}^{(j)2}}}\frac{1}{m+x},  \label{jcnsemi}
\end{equation}%
with the notation%
\begin{equation}
m_{n}^{(\pm )}=\sqrt{k_{n}^{(\pm )2}+m^{2}},\;k_{n}^{(\pm )}=2\pi (n+\Phi
/\Phi _{0}\pm \alpha )/L.  \label{mn}
\end{equation}%
In Eq. (\ref{jcnsemi}), $z=0$ corresponds to the edge of the nanotube. An
alternative expression is obtained by using Eq. (\ref{j15}):%
\begin{eqnarray}
\langle j^{2}\rangle _{\mathrm{(cn)}}^{(1)} &=&-\frac{mv_{F}}{\pi ^{2}}%
\sum_{j=+,-}\sin (2\pi (\Phi /\Phi _{0}+j\alpha ))\int_{0}^{\infty }dx\,x
\notag \\
&&\times \lbrack \cosh (L\sqrt{x^{2}+m})-\cos (2\pi (\Phi /\Phi _{0}+j\alpha
))]^{-1}  \notag \\
&&\times \int_{0}^{1}dy\frac{m\cos (2zxy)-xy\sin (2zxy)}{(m^{2}+x^{2}y^{2})%
\sqrt{1-y^{2}}}.  \label{jcnsemi2}
\end{eqnarray}%
This expression is more convenient for the evaluation of the
current near the edge and for large values of the nanotube
diameter. The asymptotic expressions in various limiting cases are
directly obtained from those for general $D$ discussed in the
previous section. The electric current corresponding to the VEV of
the fermionic current discussed in this section is of the order
$|e|v_{F}/L$. The persistent currents in normal metal rings with
this order of magnitude have been recently detected in Ref.
\cite{Bluh09}.

\begin{figure}[tbph]
\begin{center}
\begin{tabular}{cc}
\epsfig{figure=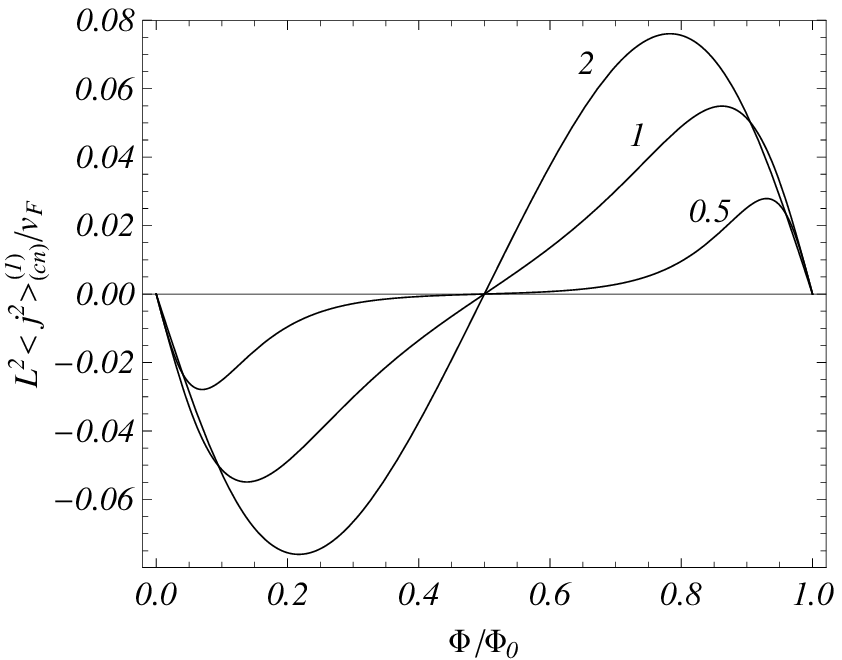,width=7.cm,height=6.cm} & \quad %
\epsfig{figure=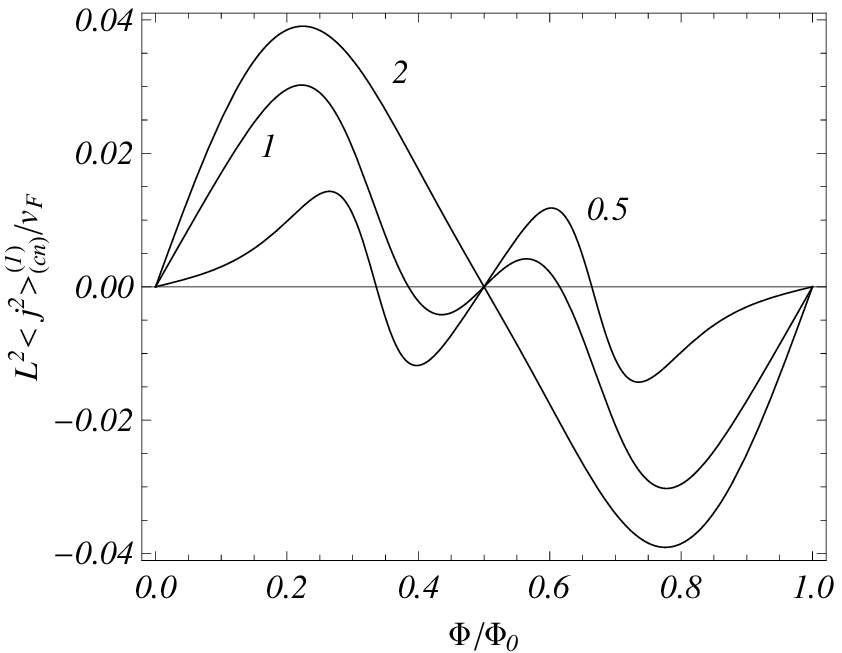,width=7.cm,height=6cm}%
\end{tabular}%
\end{center}
\caption{Edge induced part in the fermionic current for
semi-infinite metallic (left panel) and semiconducting (right
panel) carbon nanotubes as a function of the magnetic flux. The
graphs are plotted for $mz=0.5$ and for different values of the
parameter $mL$ (numbers near the curves).} \label{fig3}
\end{figure}

In Fig. \ref{fig3} we display the dependence of $L^{2}\langle j^{2}\rangle _{%
\mathrm{(cn)}}^{(1)}$ on the magnetic flux, in units of the flux quantum, for $%
mz=0.5$. The numbers near the curves correspond to the values of
$mL$. The left/right panel correspond to metallic/semiconducting
nanotubes.

In the case of a nanotube of finite length, $a$, the general expression (\ref%
{jpar3}) takes the form%
\begin{eqnarray}
&& \langle j^{2}\rangle _{\mathrm{(cn)}} =\langle j^{2}\rangle _{\mathrm{(cn)}%
}^{(0)}-\frac{2v_{F}}{\pi L}\sum_{n=-\infty }^{+\infty
}\sum_{j=+,-}k_{n}^{(j)}\,  \notag \\
&& \qquad \times \int_{m_{n}^{(j)}}^{\infty }dx\frac{1+\frac{m}{x-m}e^{ax}\cosh %
\left[ (a-2z)x\right] }{\left( \frac{x+m}{x-m}e^{2ax}+1\right) \sqrt{%
x^{2}-m_{n}^{(j)2}}}.  \label{jcn}
\end{eqnarray}%
In the absence of a magnetic flux, the sublattices give opposite
contributions to the fermionic current, and the latter vanishes.
For a massless field one finds%
\begin{equation}
\langle j^{2}\rangle _{\mathrm{(cn)}}=\langle j^{2}\rangle _{\mathrm{(cn)}%
}^{(0)}+\frac{2v_{F}}{\pi L}\sum_{n=-\infty }^{+\infty
}\sum_{j=+,-}k_{n}^{(j)}\sum_{s=1}^{\infty }(-1)^{s}K_{0}(2sa|k_{n}^{(j)}|).
\label{jcnm0}
\end{equation}%
As already mentioned, the fermionic current is a periodic function of the
magnetic flux, with period equal to the flux quantum $\Phi _{0}$.

\begin{figure}[tbph]
\begin{center}
\begin{tabular}{cc}
\epsfig{figure=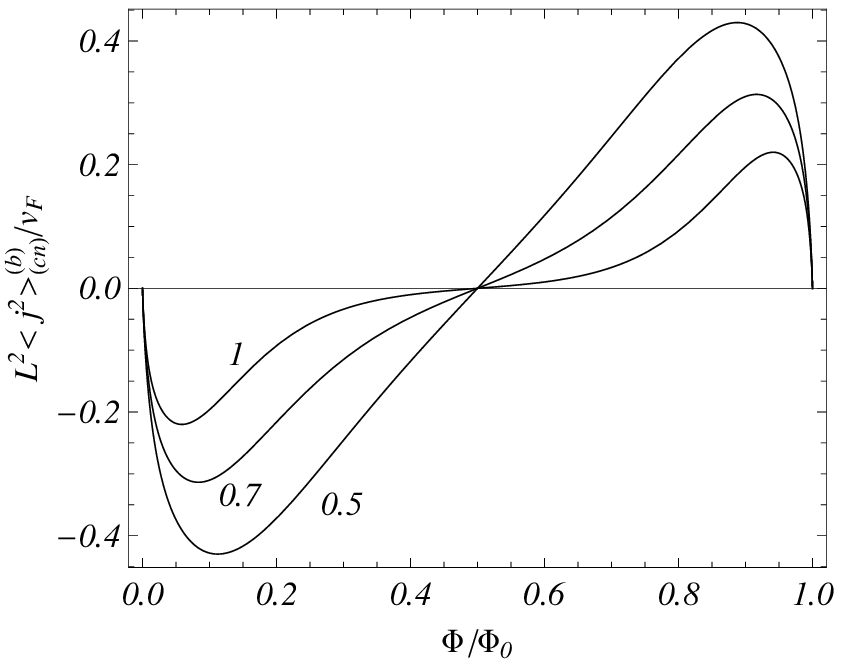,width=7.cm,height=6.cm} & \quad %
\epsfig{figure=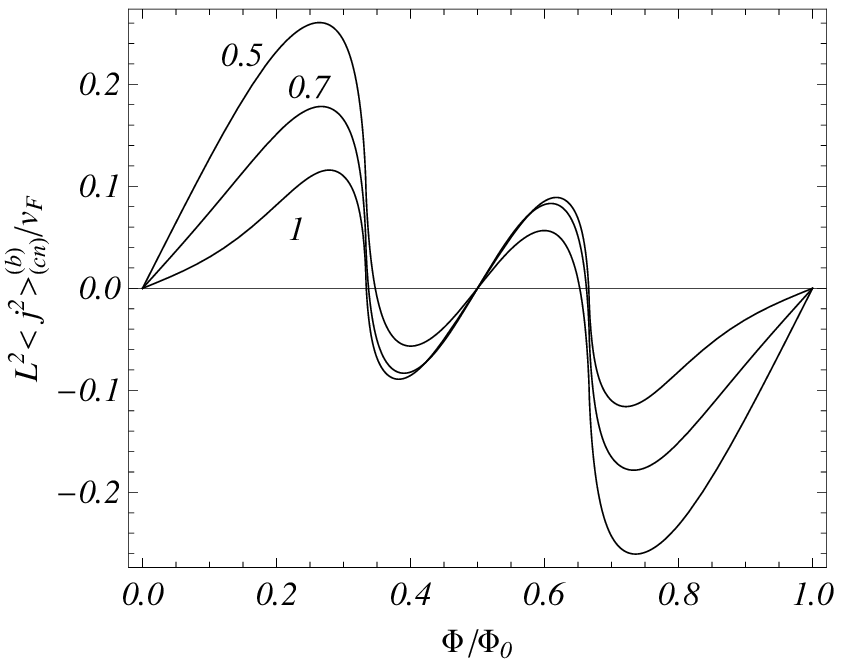,width=7.cm,height=6.cm}%
\end{tabular}%
\end{center}
\caption{Boundary induced part in the VEV of the fermionic current
for a massless field in finite-length metallic (left plot) and
semiconducting (right plot) nanotubes, as a function of the
magnetic flux. Numbers near the curves correspond to the values of
the parameter $a/L$.} \label{fig4}
\end{figure}

In Fig.~\ref{fig4} we have plotted the boundary induced part, $\langle
j^{2}\rangle _{\mathrm{(cn)}}^{\mathrm{(b)}}=\langle j^{2}\rangle _{\mathrm{%
(cn)}}-\langle j^{2}\rangle _{\mathrm{(cn)}}^{(0)}$, for a
massless field in metallic (left panel) and semiconducting (right
panel) finite-length nanotubes, as a function of the magnetic
flux. Numbers near the curves correspond to the values of the
parameter $a/L$.

\begin{figure}[tbph]
\begin{center}
\begin{tabular}{cc}
\epsfig{figure=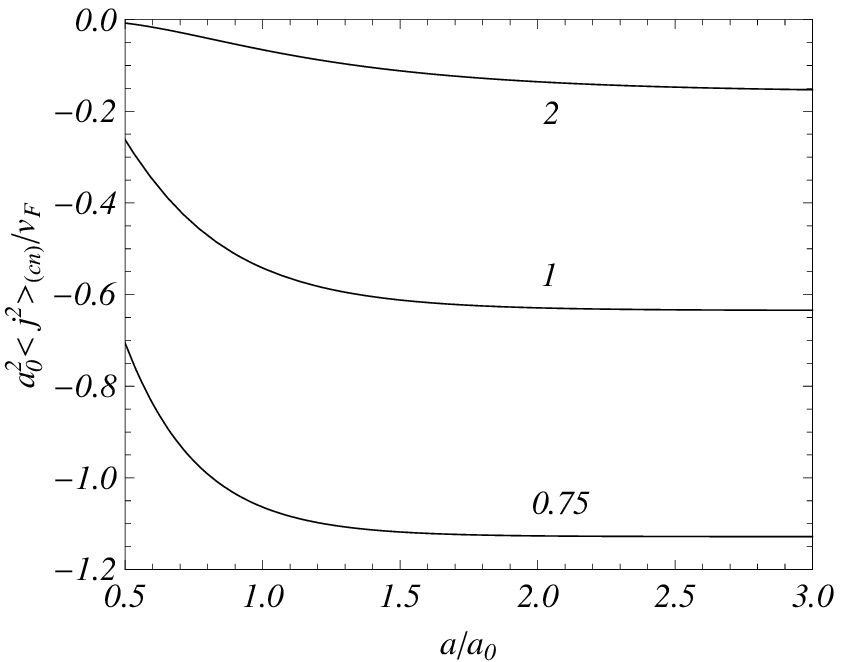,width=7.cm,height=6.cm} & \quad %
\epsfig{figure=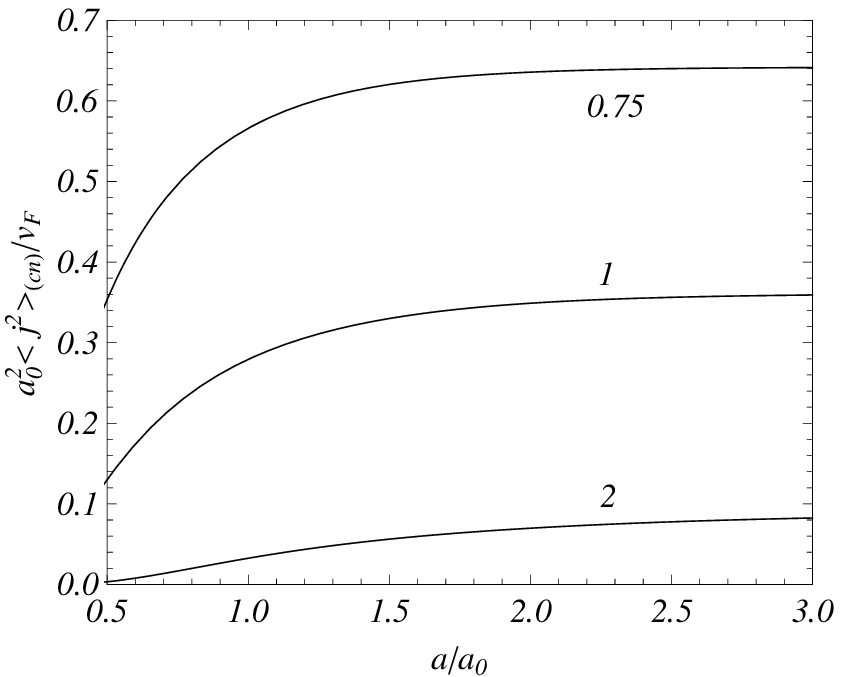,width=7.cm,height=6.cm}%
\end{tabular}%
\end{center}
\caption{VEV of the fermionic current for a massless field in
metallic (left plot) and semiconducting (right plot) nanotubes, as
a function of the tube length. The graphs are plotted for $\Phi
=0.8\Phi _{0}$ and the numbers near the curves correspond to the
values of the parameter $L/a_{0}$.} \label{fig5}
\end{figure}

The dependence of the current density on the length of the
nanotube is displayed in Fig. \ref{fig5}, in the model with a
massless fermionic field. The numbers near the curves correspond
to the length of the compact dimension in units of a fixed length
scale $a_{0}$, and for the magnetic flux we have taken the value
$\Phi =0.8\Phi _{0}$. The left and right panels correspond to
metallic and semiconducting nanotubes. As it is seen, for long
nanotubes the edge induced effects are small and the current tends
the corresponding value for an infinite length nanotube.

\section{Conclusion}

\label{sec:Conc}

We have investigated combined effects of compact spatial
dimensions and boundaries on the VEV of the fermionic current. The
geometry of boundaries is given by two parallel plates on which
the Dirac fermion field obeys MIT bag boundary conditions.
Imposing these conditions in the region between the plates leads
to the discrete eigenvalues for the component of the momentum
perpendicular to the plates. For the case of a massive field, they
are solutions of the transcendental equation (\ref{NormMod}).
Along the compact dimensions, the periodicity conditions
(\ref{BC}) are imposed with constant phases $\alpha _{l}$. In
addition, we assume the presence of a constant gauge field which
leads to the Aharonov-Bohm-like effect on the VEV
of the fermionic current. The application of the summation formula (\ref%
{Abel-Plan}) allowed us to decompose the VEV as the sum of topological,
single plate and interference parts. In this way the renormalization is
reduced to the one for the boundary-free geometry.

Similar to the case of the geometry without boundaries, the VEV of
the charge density vanishes. As for what concerns the spatial
components of the current density, only the components along
compact dimensions are non-vanishing. The VEV of the corresponding
fermionic current depends on the phases in the periodicity
conditions and on the components of the vector potential in the
combination (\ref{alftilde}). The VEV\ is a periodic odd function
of this
parameter with the period equal to 1. It vanishes for special cases $\tilde{%
\alpha}_{l}=0$ and $\tilde{\alpha}_{l}=1/2$. This is the case for untwisted
and twisted fermion fields in the absence of the gauge field.

Firstly, we have considered the geometry of a single plate. The
boundary induced part in the VEV for the component of the
fermionic current along the $l$-th compact dimension is given by
Eq. (\ref{j12}). This part vanishes for a massless field. An
alternative expression, Eq. (\ref{j15}), is derived in Appendix
\ref{sec:Appendix}, by using the summation formula
(\ref{SumForm}). Unlike the case of the fermionic condensate and
the VEV of the energy-momentum tensor, the VEV\ of the fermionic
current is finite on the boundary. In both limits of small and
large values of the compact dimension length, the boundary induced
part vanishes. At large distances from the plate, the asymptotic
behavior of the fermionic current essentially depends on the
phases present in the periodicity conditions along the compact
dimensions.

In the region between two plates, the VEV of the fermionic current is
presented in the form (\ref{jpar3}), where the first term on the right-hand
side is the corresponding VEV in the boundary-free space with topology \ $%
R^{p+1}\times (S^{1})^{q}$. The second term is induced by the boundaries. An
alternative representation is given by Eq. (\ref{jinterf}) with the
interference term defined by Eq. (\ref{Deltj}). For a massless field the
distribution of the boundary induced part is uniform and is presented in two
equivalent forms, Eqs. (\ref{jparm0}) and (\ref{jparm0B}). At large
separations between the plates the boundary induced part is exponentially
small (see Eq. (\ref{jlLargSepm})).

In the massless case the Dirac equation is conformally invariant
for all spacetime dimensions. By taking into account that the bag
boundary condition also is conformally invariant, and using the
expression given above, we can generate the VEV of the fermionic
current in conformally-flat spacetimes with compact spatial
dimensions in the presence of boundaries. The corresponding VEV is
given by Eq.~(\ref{jCurv}), where the first term in the right-hand
side is the VEV for a conformally-flat spacetime with trivial
topology, in the absence of boundaries, and the second term is
induced by the nontrivial topology and by the boundaries.
Important special cases correspond to background de Sitter and
anti-de Sitter spacetimes. As an example, we have considered the
fermionic current induced on the visible brane in
higher-dimensional generalizations of Randall-Sundrum-type
braneworlds with internal spaces.

As an application of the general formulas, in
Sect.~\ref{sec:Nano}, we have considered the VEV of the fermionic
current in finite-length carbon nanotubes, described within the
framework of the Dirac-like theory for the electronic subsystem.
Recently a number of techniques allowed one to synthesize
ultra-short carbon nanotubes which are especially useful in
medical applications. For this type of nanotubes, the tube length
can be of the order of magnitude of the compact dimension length
(this is especially the case for the outermost tubes of
multi-walled nanotubes) and the finite length effects may be
important. The corresponding expressions for the VEV\ of the
fermionic current are obtained by summing the contributions of the
two triangular sublattices of the graphene lattice, with opposite
signs of the phases along the compact dimension. Due to the
cancellation between these sublattices, the current vanishes in
the absence of a magnetic flux through the cross section of the
nanotube. A magnetic flux through the cross section of the
nanotube will break the symmetry allowing the current to flow
along the compact dimension. The current is a periodic function of
the flux with the period of the flux quantum $\Phi _{0}$. The pure
topological contribution is then given by Eq. (\ref{jrs}), where the values $%
\alpha =0$ and $\alpha =1/3$ correspond to metallic and to semiconducting
nanotubes, respectively. For a semi-infinite nanotube, the part induced by
the edge is presented in two equivalent forms, Eqs. (\ref{jcnsemi}) and (\ref%
{jcnsemi2}). For a finite-length nanotube the VEV of the current is given by
Eqs. (\ref{jcn}) and (\ref{jcnm0}) for massive and massless fields,
respectively. For long nanotubes the interference effects between the edges
are small and the single edge parts dominate.

\section*{Acknowledgments}

A.A.S. gratefully acknowledges the hospitality of the INFN, Laboratori
Nazionali di Frascati (Frascati, Italy) where part of this work was done.

\appendix

\section{Alternative representation for a single plate induced part}

\label{sec:Appendix}

An alternative representation for the part in the VEV\ of the fermionic
current, induced by a single plate at $z=0$, can be obtained by applying to
the series over $n_{l}$ in Eq. (\ref{j11}) the summation formula \cite%
{Bell09b,Beze08}
\begin{eqnarray}
&& \sum_{n_{l}=-\infty }^{\infty }g(n_{l}+\tilde{\alpha}_{l})f(\left\vert n_{l}+%
\tilde{\alpha}_{l}\right\vert ) =\int_{0}^{\infty
}du[g(u)+g(-u)]f(u)
\notag \\
&& \qquad +i\int_{0}^{\infty }du\,[f(iu)-f(-iu)]\sum_{\lambda =\pm 1}\frac{%
g(i\lambda u)}{e^{2\pi (u+i\lambda \tilde{\alpha}_{l})}-1}.  \label{SumForm}
\end{eqnarray}%
For the corresponding series in Eq. (\ref{j11}) one has $g(u)=u$ and the
first integral in the right-hand side of this formula vanishes. By
integrating over $\mathbf{k}_{p}$ with the help of the analog of Eq. (\ref%
{IntRel2}), the VEV is presented in the form%
\begin{eqnarray}
\langle j^{l}\rangle ^{(1)} &=&-\frac{N_{D}A_{p}}{2\pi V_{q}}mL_{l}\sin
(2\pi \tilde{\alpha}_{l})\sum_{\mathbf{n}_{q-1}\in \mathbf{Z}%
^{q-1}}\int_{0}^{\infty }dx\,\left( \frac{e^{2ixz}}{m-ix}+\frac{e^{-2ixz}}{%
m+ix}\right)   \notag \\
&&\times \int_{\sqrt{x^{2}+m_{\mathbf{n}_{q-1}}^{2}}}^{\infty }dy\,y\frac{%
(y^{2}-x^{2}-m_{\mathbf{n}_{q-1}}^{2})^{(p-1)/2}}{\cosh (yL_{l})-\cos (2\pi
\tilde{\alpha}_{l})},  \label{j14}
\end{eqnarray}%
with the notation%
\begin{equation}
m_{\mathbf{n}_{q-1}}^{2}=\sum_{i=p+2,i\neq l}^{D}[2\pi (n_{i}+\tilde{\alpha}%
_{i})/L_{i}]^{2}+m^{2}.  \label{mnq-1}
\end{equation}%
Next, introducing instead of $y$ a new integration variable $t=\sqrt{%
y^{2}-x^{2}-m_{\mathbf{n}_{q-1}}^{2}}$, we pass to polar coordinates in the
plane $(x,t)$. After some additional transformations, the expression for the
VEV of the fermionic current reads

\begin{eqnarray}
\langle j^{l}\rangle ^{(1)} &=&-\frac{N_{D}A_{p}}{\pi V_{q}}mL_{l}\sin (2\pi
\tilde{\alpha}_{l})\sum_{\mathbf{n}_{q-1}\in \mathbf{Z}^{q-1}}\int_{0}^{%
\infty }dx\,x^{p+1}  \notag \\
&&\times \lbrack \cosh (L_{l}\sqrt{x^{2}+m_{\mathbf{n}_{q-1}}^{2}})-\cos
(2\pi \tilde{\alpha}_{l})]^{-1}  \notag \\
&&\times \int_{0}^{1}dy\frac{m\cos (2zxy)-xy\sin (2zxy)}{%
(m^{2}+x^{2}y^{2})(1-y^{2})^{(1-p)/2}}.  \label{j15}
\end{eqnarray}%
Note that in a model with a single compact dimension one has $m_{\mathbf{n}%
_{q-1}}=m$ and the summation in Eq. (\ref{j15}) is absent. By numerical
calculations we have checked the equivalence of Eqs. (\ref{j12}) and (\ref%
{j15}).


\begin{thebibliography}{99}
\bibitem{Most97} E. Elizalde, S.D. Odintsov, A. Romeo, A.A. Bytsenko and S.
Zerbini, \textit{Zeta Regularization Techniques with Applications} (World
Scientific, Singapore, 1994); V.M. Mostepanenko and N.N. Trunov, \textit{The
Casimir Effect and Its Applications} (Clarendon, Oxford, 1997); K.A. Milton,
\textit{The Casimir Effect: Physical Manifestation of Zero-Point Energy}
(World Scientific, Singapore, 2002); M. Bordag, G.L. Klimchitskaya, U.
Mohideen, and V.M. Mostepanenko, \textit{Advances in the Casimir Effect}
(Oxford University Press, Oxford, 2009); \textit{Lecture Notes in Physics:
Casimir Physics,} Vol. 834, edited by D. Dalvit, P. Milonni, D. Roberts, and
F. da Rosa (Springer, Berlin, 2011).

\bibitem{Lind04n} A. Linde, JCAP \textbf{0410}, 004 (2004).

\bibitem{Cast09} A.H. Castro Neto, F. Guinea, N.M.R. Peres, K.S. Novoselov,
and A.K. Geim, Rev. Mod. Phys. \textbf{81}, 109 (2009).

\bibitem{Eliz01} E. Elizalde, Phys. Lett. B \textbf{516}, 143 (2001); C.L.
Gardner, Phys. Lett. B \textbf{524}, 21 (2002); K.A. Milton, Grav. Cosmol.
\textbf{9}, 66 (2003); E. Elizalde, J. Phys. A \textbf{39}, 6299 (2006); B.
Green and J. Levin, JHEP \textbf{0711}, 096 (2007); P. Burikham, A.
Chatrabhuti, P. Patcharamaneepakorn, and K. Pimsamarn, JHEP \textbf{0807},
013 (2008).

\bibitem{Most87} G.L. Klimchitskaya, U. Mohidden, and V.M. Mostepanenko,
Rev. Mod. Phys. \textbf{81}, 1827 (2009).

\bibitem{Chen06} H.B. Cheng, Phys. Lett. B \textbf{643}, 311 (2006); H.B.
Cheng, Phys. Lett. B \textbf{668}, 72 (2008); S.A. Fulling and K. Kirsten,
Phys. Lett. B \textbf{671}, 179 (2009); K. Kirsten and S.A. Fulling, Phys.
Rev. D \textbf{79}, 065019 (2009); E. Elizalde, S.D. Odintsov, and A.A.
Saharian, Phys. Rev. D \textbf{79}, 065023 (2009); L.P. Teo, Phys. Lett. B
\textbf{672}, 190 (2009); L.P. Teo, Nucl. Phys. B \textbf{819}, 431 (2009);
L.P. Teo, JHEP \textbf{0906}, 076 (2009); L.P. Teo, JHEP \textbf{0911}, 095
(2009).

\bibitem{Popp04} K. Poppenhaeger, S. Hossenfelder, S. Hofmann, and M.
Bleicher, Phys. Lett. B \textbf{582}, 1 (2004); A. Edery and V.N.
Marachevsky, Phys. Rev. D \textbf{78}, 025021 (2008); A. Edery and V.N.
Marachevsky, JHEP \textbf{0812}, 035 (2008); F. Pascoal, L.F.A. Oliveira,
F.S.S. Rosa, and C. Farina, Braz. J. Phys. \textbf{38}, 581 (2008); L.
Perivolaropoulos, Phys. Rev. D \textbf{77}, 107301 (2008).

\bibitem{Bell09b} S. Bellucci and A.A. Saharian, Phys. Rev. D \textbf{79},
085019 (2009); S. Bellucci and A.A. Saharian, Phys. Rev. D \textbf{80},
105003 (2009).

\bibitem{Eliz11} E. Elizalde, S.D. Odintsov, and A.A. Saharian, Phys. Rev. D
\textbf{83}, 105023 (2011).

\bibitem{Flac03} A. Flachi, J. Garriga, O. Pujol\`{a}s, and T. Tanaka, J.
High Energy Phys. \textbf{0308}, 053 (2003); A. Flachi and O. Pujol\`{a}s,
Phys. Rev. D \textbf{68}, 025023 (2003); A.A. Saharian, Phys. Rev. D \textbf{%
73}, 044012 (2006); A.A. Saharian, Phys. Rev. D \textbf{73}, 064019 (2006);
A.A. Saharian, Phys. Rev. D \textbf{74}, 124009 (2006); R. Linares, H.A.
Morales-T\'{e}cotl, and O. Pedraza, Phys. Rev. D \textbf{77}, 066012 (2008);
M. Frank, N. Saad, and I. Turan, Phys. Rev. D \textbf{78}, 055014 (2008).

\bibitem{Gold00} S. Nojiri, S.D. Odintsov, and S. Zerbini, Classical Quantum
Gravity \textbf{17}, 4855 (2000); W. Goldberger and I. Rothstein, Phys.
Lett. B \textbf{491}, 339 (2000); A. Flachi and D. J. Toms, Nucl. Phys. B
\textbf{610}, 144 (2001); J. Garriga, O. Pujol\`{a}s, and T. Tanaka, Nucl.
Phys. B \textbf{605}, 192 (2001); E. Elizalde, S. Nojiri, S.D. Odintsov, and
S. Ogushi, Phys. Rev. D \textbf{67}, 063515 (2003); A.A. Saharian and M.R.
Setare, Phys. Lett. B \textbf{584}, 306 (2004); A. Knapman and D. J. Toms,
Phys. Rev. D \textbf{69}, 044023 (2004); A. Flachi, A. Knapman, W. Naylor,
and M. Sasaki, Phys. Rev. D \textbf{70}, 124011 (2004); A. A. Saharian,
Nucl. Phys. B \textbf{712}, 196 (2005); M. Frank, I. Turan, and L. Ziegler,
Phys. Rev. D \textbf{76}, 015008 (2007); L.P. Teo, Phys. Lett. B \textbf{682}%
, 259 (2009); A. Flachi and T. Tanaka, Phys. Rev. D \textbf{80}, 124022
(2009); L.P. Teo, Phys. Rev. D \textbf{82}, 027902 (2010); L.P. Teo, JHEP
\textbf{1010}, 019 (2010).

\bibitem{Bell10} S. Bellucci, A.A. Saharian, and V.M. Bardeghyan, Phys. Rev.
D \textbf{82}, 065011 (2010).

\bibitem{Beze10} E.R. Bezerra de Mello, V.B. Bezerra, A.A. Saharian, and
V.M. Bardeghyan, Phys. Rev. D \textbf{82}, 085033 (2010).

\bibitem{Rome02} A. Romeo and A. A. Saharian, J. Phys. A: Math. Gen. \textbf{%
35}, 1297 (2002).

\bibitem{Saha08Rev} A. A. Saharian, \textit{The Generalized Abel-Plana
Formula with Applications to Bessel Functions and Casimir Effect} (Yerevan
State University Publishing House, Yerevan, 2008); Preprint ICTP/2007/082;
arXiv:0708.1187.

\bibitem{Birr82} N.D. Birrell and P.C.W. Davis, \textit{Quantum Fields in
Curved Space} (Cambridge University Press, Cambridge, England, 1982).

\bibitem{Sait98} R. Saito, G. Dresselhaus, and M. S. Dresselhaus, \textit{%
Physical Properties of Carbon Nanotubes} (Imperial College Press,
London, 1998); S. Bellucci, Phys. Stat. Sol. (c) \textbf{2}, 34
(2005); S. Bellucci, Nucl. Instr. and Meth. B \textbf{234}, 57
(2005); C. Dupas, P. Houdy, and M. Lahmani (Editors),
\textit{Nanoscience: Nanotechnologies and Nanophysics} (Springer,
Berlin, 2007); J.-C. Charlier, X. Blase, and S. Roche, Rev. Mod.
Phys. \textbf{79}, 677 (2007).

\bibitem{Gusy95} V.P. Gusynin, V.A. Miransky, and I.A. Shovkovy, Phys. Rev.
D \textbf{52}, 4718 (1995); C. Chamon, Phys. Rev. B \textbf{62}, 2806
(2000); C.-Y. Hou, C. Chamon, and C. Mudry, Phys. Rev. Lett. \textbf{98},
186809 (2007); G. Giovannetti, P.A. Khomyakov, G. Brocks, P.J. Kelly, and J.
van den Brink, Phys. Rev. B \textbf{76}, 073103 (2007); S.Y. Zhou et al.,
Nature Mater. \textbf{6}, 770 (2007); G.W. Semenoff, V. Semenoff, and F.
Zhou, Phys. Rev. Lett. \textbf{101}, 087204 (2008); D. Haberer et al., Nano
Lett. \textbf{10}, 3366 (2010); R. Balog et al., Nature Mater. \textbf{9},
315 (2010).

\bibitem{Ashc06} Z. Gu et al., Nano Letters \textbf{2 (9)}, 1009 (2002);
J.M. Ashcroft et al., Nanotechnology \textbf{17}, 5033 (2006).

\bibitem{Bord06} M. Bordag, B. Geyer, G.L. Klimchitskaya, and V.M.
Mostepanenko, Phys. Rev. B \textbf{74}, 205431 (2006); M. Bordag, I.V.
Fialkovsky, D.M. Gitman, and D.V. Vassilevich, Phys. Rev. B \textbf{80},
245406 (2009); G. G\'{o}mez-Santos, Phys. Rev. B \textbf{80}, 245424 (2009);
I.V. Fialkovsky, V.N. Marachevsky, and D.V. Vassilevich, Phys. Rev. B
\textbf{84}, 035446 (2011); D. Drosdoff and L.M. Woods, Phys. Rev. A \textbf{%
84}, 062501 (2011); B.E. Sernelius, Europhys. Lett. \textbf{9}5, 57003
(2011); V. Svetovoy, Z. Moktadyr, M. Elwenspoek, and H. Mizuta,
arXiv:1108.3856; D. Drosdoff et al., arXiv:1204.4438.

\bibitem{Chur10} Yu.V. Churkin, A.B. Fedortsov, G.L. Klimchitskaya, and V.A.
Yurova, Phys. Rev. B \textbf{82}, 165433 (2010); M. Chaichian, G.L.
Klimchitskaya, V.M. Mostepanenko, and A. Tureanu, arXiv:1207.3788.

\bibitem{Bluh09} H. Bluhm et al., Phys. Rev. Lett. \textbf{102}, 136802
(2009); A.C. Bleszynski-Jayich et al., Science \textbf{326}, 272
(2009).

\bibitem{Beze08} E.R. Bezerra de Mello and A.A. Saharian, Phys. Rev. D
\textbf{78}, 045021 (2008).
\end{thebibliography}
\end{document}